\documentclass[twocolumn]{aastex62}

\usepackage{graphicx}	% Including figure files
\usepackage{amsmath}	% Advanced maths commands
\usepackage{amssymb}	% Extra maths symbols

\usepackage{bm}
\usepackage[shortlabels]{enumitem}

\received{January 1, 2018}
\revised{January 7, 2018}
\accepted{\today}
\submitjournal{ApJ}

\shorttitle{magnetic Kelvin--Helmholtz mixing}
\shortauthors{A. Hillier \& I. Arregui}
\begin{document}

\title{Coronal cooling as a result of mixing by the nonlinear Kelvin--Helmholtz instability}

\correspondingauthor{Andrew Hillier}
\email{a.s.hillier@exeter.ac.uk}

\author[0000-0002-0851-5362]{Andrew Hillier}
\affil{CEMPS\\ University of Exeter\\ Exeter EX4 4QF U.K.}

\author[0000-0002-7008-7661]{Inigo Arregui}
\affiliation{Instituto de Astrof\'{i}sica de Canarias\\ V\'{i}a L\'{a}ctea s/n\\ E-38205 La Laguna, Tenerife, Spain}
\affiliation{Departamento de Astrof\'{i}sica\\ Universidad de La Laguna\\ E-38206 La Laguna, Tenerife, Spain}



\begin{abstract}

Recent observations show cool, oscillating prominence threads fading when observed in cool spectral lines and appearing in warm spectral lines.
A proposed mechanism to explain the observed temperature evolution is that the threads were heated by turbulence driven by the Kelvin--Helmholtz instability that {developed as} a result of wave-driven shear flows on the surface of the thread.
As the Kelvin--Helmholtz instability is an instability that works to mix the two fluids either side of the velocity shear layer, in the solar corona it can be expected to work by mixing the cool prominence material with that of the hot corona to form a warm boundary layer.
In this paper we develop a simple phenomenological model of nonlinear Kelvin--Helmholtz mixing, using it to determine the characteristic density and temperature of the mixing layer, which for the case under study with constant pressure across the two fluids are $\rho_{\rm mixed}=\sqrt{\rho_1\rho_2}$ and $T_{\rm mixed}=\sqrt{T_1T_2}$.
One result from the model is that it provides an accurate, as determined by comparison with simulation results, determination of the kinetic energy in the mean velocity field.
A consequence of this is that the magnitude of turbulence, and with it the energy that can be dissipated on fast time-scales, as driven by this instability can be determined.
For the prominence-corona system, the mean temperature rise possible from turbulent heating is estimated to be less than 1\% of the characteristic temperature (which is found to be $T_{\rm mixed}=10^5$\,K).
These results highlight that mixing, and not heating, are likely to be the cause of the observed transition between cool to warm material in \citet{OKA2015}.
One consequence of this result is that the mixing creates a region with higher radiative loss rates {on average} than either of the original fluids, meaning that this instability could contribute a net loss of thermal energy from the corona, i.e. coronal cooling. 

\end{abstract}



\section{Introduction}\label{INTRO}

The dissipation of magnetohydrodynamic (MHD) wave energy has been regarded for decades as a relevant agent in explaining the heating of the solar corona \citep[see][for a recent review]{ARREGUI2015}. 
Since first proposed by \citet{ION1978}, the resonant absorption of surface Alfv\'{e}n waves offers a means to transfer wave energy from large to small spatial scales, thus enhancing dissipative processes \citep{HOLL1978, WEN1974, WEN1978, WEN1979}. 
Theoretical and numerical advances have recently shown that the nature of the resonantly damped transverse kink wave \citep{GOOS2009} and its associated non-linear dynamics leads to the development of Kelvin--Helmholtz (KH) unstable flows \citep{TERR2008, ANT2014, ANT2015, ANT2016, ANT2017, TERR2018}. 
The instability arises in connection with resonant absorption processes because of the creation of a shear velocity pattern around the resonance, but owes its existence to the presence of a discontinuous shear flow even in models with a density jump at the boundary of the waveguide. 
It operates by extracting energy from the large scale dynamics to spread it among different spatial scales and locations. 
The cause of the heating, though, is still under investigation \citep{MAGYAR2016, HOW2017,KARA2017, ANT2018}, but as well as heating in some cases it has been shown that mixing plays a dominant role in the thermal evolution \citep[e.g.][]{MAGYAR2016,KARA2017}.

Obtaining observational evidence about these small scale physical processes is being pursued only recently. Prominence plasmas offer a natural laboratory in this context, because of the occurrence of complex oscillatory and flow patterns at both larges scales \citep{BERG2008, HILL2013} and in their fine structure \citep{OKA2007,OKA2015}. 
In particular, recent observations by \citet{OKA2015} using the Interface Region Imaging Spectrograph \citep[IRIS;][]{DEPON2014} and Hinode Solar Optical Telescope \citep[SOT;][]{TSU2008} found oscillations of prominences threads that display velocity features consistent with resonant absorption. 
These threads were often found to fade from the cool passbands on \ion{Mg}{2} K (observed by IRIS) and \ion{Ca}{2} H (observed by Hinode SOT), whilst becoming brighter in warmer (\ion{Si}{4}) IRIS passbands.
\citet{ANT2015} simulated these processes, showing that the concentration of the wave energy onto the surface of the flux tube produced shear flows large enough to develop an instability. 
The key process they proposed to be behind the observed temperature evolution was heating as a result of turbulence driven by the magnetic Kelvin--Helmholtz instability.
This mechanism has also been found in simulations of oscillating coronal loops \citep{TERR2008}.

The KH instability breaks up shear flows by  creating vorticies at the shear layer \citep{CHAN1961}, mixing the two regions together. These vorticies may themselves break up into turbulence via secondary 3D instabilities. {With the inclusion of magnetic fields, magnetic tension works to suppress the KH instability. Therefore, the most unstable modes become those that vary little along the field.}
This instability has been found to occur in a number of different situations in the solar atmosphere including in the interaction between prominences and bubbles that form below them \citep{RYU2010,BERG2010,BERG2017}, or internal prominence motions \citep{HILL2018, YANG2018}, and as a result of eruptions in the solar atmosphere \citep{FOU2011, OFMAN2011, MOSTL2013}.
\citet{SOLER2010} investigated how this instability develops on the surface of a rotating flux tube, a model used because of its geometrical connection to coronal loops, finding that the fundamental physics of the linear instability are not greatly altered by the change in geometry. The studies of \citet{HILL2019} and \citet{BARB2019} highlighted the important role the oscillatory nature of the flow in wave-driven KHi could have in determining stability, with both the KHi and resonance-induced parametric instabilities existing.
Once the linear instability has developed, nonlinearities form and it is in this nonlinear stage that the important processes of heating and mixing are driven.

The nonlinear stage of the instability is where the key dynamics to explain the observations should be occurring.
\citet{RYU2000} investigated the 3D evolution of the MHD KHi, finding the KHi vorticies could become disrupted resulting in highly turbulent and effficient mixing of the two layers. % They found that the vortex disruption would occur if the Alfv\'{e}n Mach number of the flow dropped below unity anywhere.% This low Alfv\'{e}nic Mach number showed that the magnetic field had been highly stretched and distorted by the flow. 
A detailed analysis of the disruption process in 2D of a KHi vortex via magnetic reconnection was presented in \citet{MAK2017}, finding that significant disruption would occur when $M^2R{\rm m} =O(1)$ ($M$ is the Alfv\'{e}n Mach number and $R_{\rm m}$ is the magnetic Reynolds number).
%Reconnection driven by the KHi can be very important for plasma transport, for example in the Magnetosphere \citep[e.g.][]{NYK2001}. 
%
%In connection to these MHD studies, a huge amount can be understood about the nonlinear behaviour of the nonlinear MHD KHi by looking at the wealth of studies of the nonlinear HD KHi. 
%The spread of the mixing layer was found to be expanding in a self-similar fashion \citep[e.g.][]{WIN1974} where the width ($W$) was found to be $W\propto \Delta V t$ where $\Delta V$ is the velocity difference and $t$ is time.
%Modelling the evolution of the constant density fluid under the eddy viscosity assumption, gave a mean flow profile that of the {\bf finish}
%
%To understand the role of the density contrast in the development of MHD KHi turbulence, \citet{MATSU2004} and \citet{MATSU2010} performed a range of 2D simulations.
%\citet{MATSU2004} found that having a large density contrast would allow the baroclinic term (the term taking the form $\nabla\rho\times\nabla p$ in the vorticity equation) to drive an energy cascade in the 2D flow (this is not possible in cases with constant density), recovering spectra with $k^{-5/3}$.% The presence of a strong energy cascade can be very important for energy dissipation.
\citet{MATSU2010} studied the nonlinear evolution of high density-contrast MHD KHi in 2D simulations, finding that large asymmetric mixing layers were formed. This implies that for situations in the solar atmosphere both the magnetic field and density contrast can be important for determining the mixing dynamics.

There are three questions to which the answer would provide key information in understanding the role of the KH instability in heating the solar corona: 1) What is the temperature in the KH layer achieved purely from mixing, 2) How much heating can be driven by the instability, 3) On what timescales does this heating occur? 
In this paper we present a phenomenological model of a turbulent MHD Kelvin--Helmholtz mixing layer that we use to provide answers to these three questions. We also use 3D MHD simulations to confirm the predictions of the model and to highlight areas in which the model can be improved.

\section{Modelling the nonlinear Kelvin--Helmholtz instability mixing layer}\label{phem_model}

\begin{figure}
  \begin{center}
\includegraphics[width=8cm]{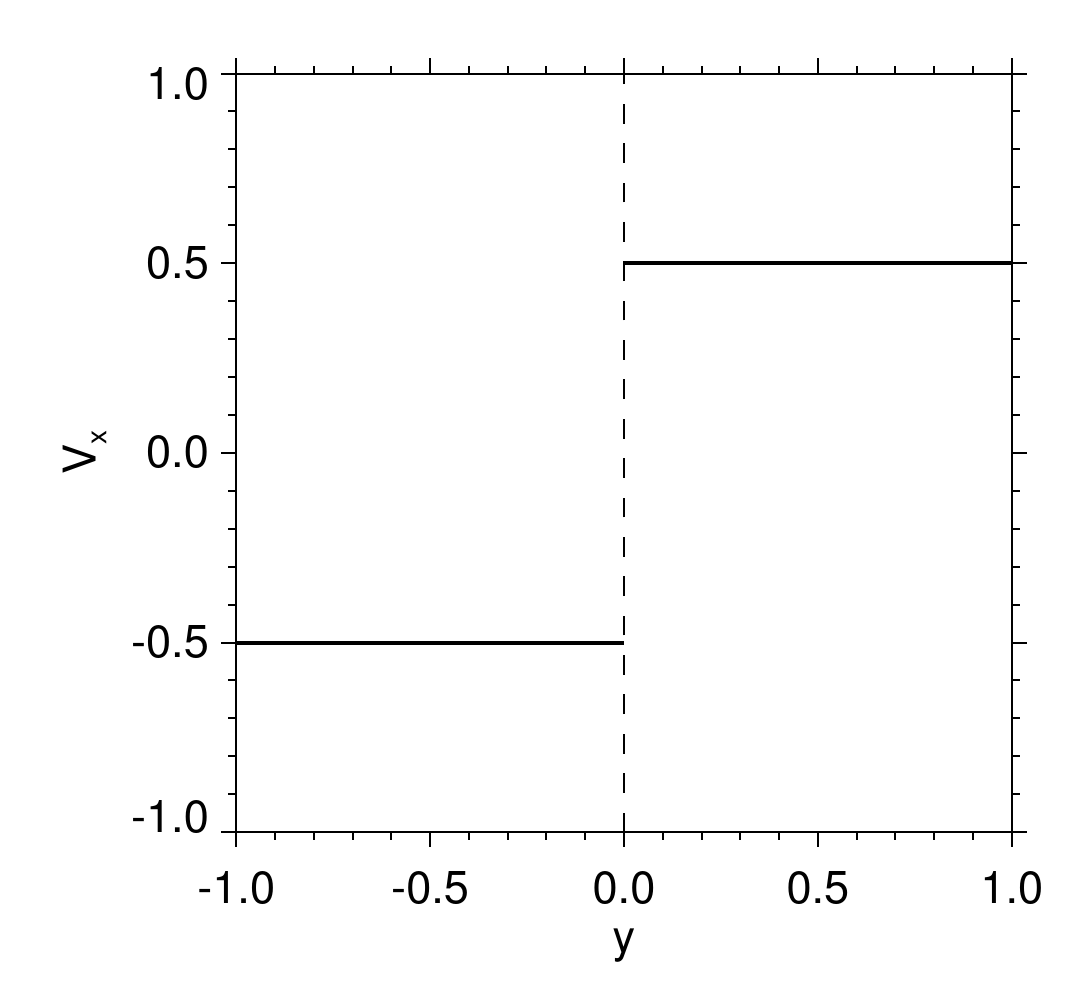}
  \end{center}
  \caption{Plot of the initial velocity distribution (normalised so that $\Delta V=1$).}
\label{vel_model}
\end{figure}

Our model aims at investigating the nonlinear solution of the magnetic KH instability as pertinent to a surface flow on an oscillating flux tube. For simplicity, a phenomenological model is considered consisting on a plane-parallel shear flow defined by:
\begin{align}
V_{x}(y)=&\begin{cases} V_1 & \mbox{if $y < 0$};\\ V_2 & \mbox{if $y > 0$}\end{cases},\\
\rho(y)=&\begin{cases} \rho_1 & \mbox{if $y < 0$};\\ \rho_2 & \mbox{if $y > 0$}\end{cases},\\
p(y)=p,\\
B_z(y)=B,
\end{align}
which is a discontinuous velocity field creating a shear layer at $y=0$ (e.g. see Fig. \ref{vel_model}) composed of two uniform layers that have their own uniform density,  but the gas pressure and magnetic field strengths are constant throughout the domain which ensures equilibrium. 
The initial velocity profile is shown in Fig. \ref{vel_model}.

\subsection{The position of the mixing layer}

For the linear instability with a discontinuity in the velocity and density at $y=0$, the instability is centred at $y=0$ with the eigenfunction decaying from there as $\exp(-k|y|)$ \citep[e.g.][]{CHAN1961}.
However, the nonlinear evolution of the instability is not required to obey the same rules.
Therefore the initial step to understanding the nonlinear mixing is to determine where would a mixing layer be centred.
Here our first assumption is introduced, we assume that the mixing works in the way that most efficiently uses the free energy associated with the initial flow.
Therefore, for a mixing layer of width across the shear layer of $2l$, it will be centred at the position $Y$ that maximises the kinetic energy associated with the shear flow.
%We propose that this point forms the centre of the mixing layer.

{As an aside, it is worth noting that generally the lengthscales of the vortical/turbulent structures in the direction of the shearflow control the width $2l$. Using as a guide the Kelvin-Stuart cat's eye vortex flow \citep{KEL1880,STU1967}, which is a steady-state flow solution producing a string of vortices given by
\begin{align}
[V_x,V_y]=V_{\infty}\frac{[\sinh(ky),\varepsilon\sin(ky)]}{\cosh(ky)+\varepsilon \cos(kx)},
%V_y=V_0\frac{\varepsilon\sin(ky)}{\cosh(ky)+\varepsilon \cos(kx)},
\end{align}
where $V_{\infty}$ is the velocity of the flow as $y\rightarrow \infty$, $k=2\pi/l_{\rm flow}$ with $l_{\rm flow}$ the length of the vortex in the direction of the shearflow, and $\varepsilon$ is the parameter which controls the localisation of the vorticity.
Based on this solution, we expect that $2l<l_{\rm flow}$ with $2l\approx l_{flow}/2$ a common ratio. As the lengthscales associated with the vortices/turbulence grow linearly with time \citep[e.g.][]{WIN1974}, the width should keep on increasing until geometric effects cause it to saturate.}

Determining the position $Y$ has one major difficulty: The kinetic energy measured for each component of a shear flow will depend on the reference frame in which the flow is being observed.
Therefore, different initial conditions would result in different layer positions. 
However, it should be expected that in situations with the same magnitude of density and velocity jump at y=0, but with different velocity values, the properties of the solution should not change, i.e. the problem is Galilean invariant.
Therefore, the question is: what is the correct reference frame to view the problem so that the kinetic energy available is the kinetic energy that can be used by the nonlinear instability. 
For this, the natural choice is to put the velocities into the zero-momentum reference frame, because it results in the removal of any mean advection in the layer.

If we take a mixing layer of width $2l$ which is centred on $Y$ where $Y$ is in the range $[-l,l]$, then a zero-momentum reference frame can be calculated for that layer.
{This is given by integrating the momentum across the layer and setting the result to zero, i.e.:
\begin{equation}
\frac{1}{\rho_1+\rho_2}\left(\int_{Y'-1}^{0} \rho_1V_1dy'+\int^{Y'+1}_{0} \rho_2V_2dy' \right)=0,
\end{equation}
where $Y'=Y/l$.
On solving this integral we have:
\begin{equation}
\alpha_1V_1(1-Y')+\alpha_2V_2(1+Y')=0,
\end{equation}
where $\alpha_{1,2}=\rho_{1,2}/(\rho_1+\rho_2)$}
Using the identity $\Delta V\equiv V_1-V_2$ we can define $V_1$ and $V_2$ as:
\begin{align}
V_1=&-\frac{\alpha_2\Delta V(1+Y')}{Y'(\alpha_1-\alpha_2)-1},\\
V_2=&\frac{\alpha_1\Delta V(1-Y')}{Y'(\alpha_1-\alpha_2)-1}.
\end{align}

\begin{figure}
  \begin{center}
\includegraphics[width=8cm]{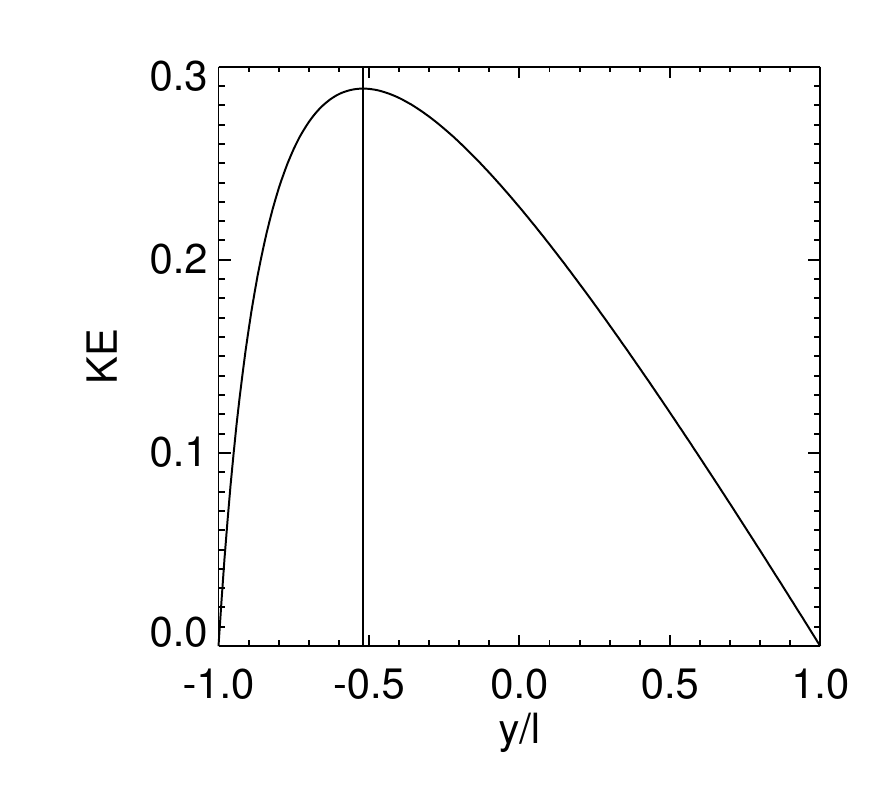}
  \end{center}
  \caption{Plot of the kinetic energy ({using the normalised density and velocity profiles}) against $Y'$. Vertical solid line gives the position of the peak value as given in Eqn \ref{y_max} }
\label{KE_pos}
\end{figure}

From this we can show that the mean kinetic energy ($KE$) for the layer in the zero-momentum frame is given by:
\begin{align}
{\rm KE}=&\rho_{\rm av}\left[ \alpha_1V_1^2\frac{1-Y'}{2}+\alpha_2V_2^2\frac{1+Y'}{2}\right]\\=&\frac{\rho_{\rm av}\alpha_1\alpha_2\Delta V^2(Y'-1)(Y'+1)}{2[Y'(\alpha_1-\alpha_2)-1]},\nonumber
\end{align}
where $\rho_{\rm av}$ is the average of the densities of the two layers.
For example the mean kinetic energy for the band from $y=-2l$ to $0$ (centred on $Y=-l$), when put in the zero-momentum frame of reference has no kinetic energy because there is no velocity shear in this region.
The same can be said for the region $y=0$ to $2l$ (centred on $Y=l$).
The mean kinetic energy peaks at somewhere between these two values at the point:
\begin{equation}\label{y_max}
Y'_{\rm max}=\frac{1-\sqrt{4\alpha_2\alpha_1}}{\alpha_1-\alpha_2}=\frac{\sqrt{\alpha_1}-\sqrt{\alpha_2}}{\sqrt{\alpha_1}+\sqrt{\alpha_2}}.%=\frac{\rho_1^{1/2}-\rho_2^{1/2}}{\rho_1^{1/2}+\rho_2^{1/2}}
\end{equation}
A key point here is that this shift is completely independent of the magnitude of the velocity shear, it is only a function of the normalised densities.
This gives the two velocities in this reference frame as:
\begin{align}
V_1=&\frac{\alpha_2\Delta V(2\alpha_1-\sqrt{4\alpha_1\alpha_2})}{\sqrt{4\alpha_1\alpha_2}(\alpha_1-\alpha_2)}=\frac{\Delta V\sqrt{\alpha_2}}{\sqrt{\alpha_1}+\sqrt{\alpha_2}}\label{left_vel}\\
V_2=&-\frac{\alpha_1\Delta V(\sqrt{4\alpha_1\alpha_2}-2\alpha_2)}{\sqrt{4\alpha_1\alpha_2}(\alpha_1-\alpha_2)}=-\frac{\Delta V\sqrt{\alpha_1}}{\sqrt{\alpha_1}+\sqrt{\alpha_2}}\label{right_vel}
\end{align}
This leads to the maximum value for the mean ${\rm KE}$ in this layer to be:
\begin{equation}\label{max_KE}
{\rm KE}_{\rm max}=\frac{\rho_{\rm av}\Delta V^2\alpha_1\alpha_2}{(\sqrt{\alpha_1}+\sqrt{\alpha_2})^2}.
\end{equation}
Figure \ref{KE_pos} shows the distribution for $KE$ for given layer positions for $\alpha_1=1/11$. The position of the peak value, as determined in Equation \ref{y_max}, is shown by the vertical line. 
One interesting property of this peak is that it is achieved when the layer is placed such that the initial kinetic energy distribution becomes continuous (i.e. the kinetic energy is the same either side of the discontinuity).

\subsection{Density, pressure, magnetic field and temperature}\label{mixed_quantities}

Once the position of the layer has been determined, the next step is to estimate the values of the average density, pressure and magnetic field in the mixed layer.
To do this, we will consider conserved quantities in MHD flows, i.e. conservation of mass, momentum, energy and magnetic flux, to determine the characteristic values of the density, pressure and magnetic field as achieved by mixing.

For a mixing layer consisting of a layer of width $2l$, the characteristic density that results from mixing ($\rho_{\rm mixed}$) can be calculated using conservation of mass from the two densities ($\rho_{1}$ and $\rho_{2}$ respectively).
That is to say that for a layer of width $2l$ centred at $Y$ the following equality must hold:
\begin{equation}
\rho_1\int_{-l+Y}^{0}dy+\rho_2\int^{l+Y}_{0}dy=\rho_{\rm mixed}\int^{l+Y}_{-l+Y}\mathrm{d}y=2l\rho_{\rm mixed}.
\end{equation}
This gives:
\begin{equation}\label{mixed_den}
\rho_{\rm mixed}=\frac{\rho_{1}(1-Y')+\rho_{2}(1+Y')}{2},
\end{equation}
where, as before, $Y'=Y/l$.
At $Y'_{\rm max}$ this equals:
\begin{equation}
\rho_{\rm mixed}=\rho_{\rm av}\sqrt{4 \alpha_1\alpha_2}=\sqrt{\rho_1\rho_2}.
\end{equation}

The mixed pressure $p_{\rm mixed}$ follows from a similar averaging process as the density.
Firstly we note that from the first law of thermodynamics we expect:
\begin{equation}
\delta U+\delta W=\delta Q,
\end{equation}
where $\delta U$ is the change in internal energy, $\delta W$ is the work done and $\delta Q$ is the heating.
In the case of no dissipation then enthalpy is conserved, i.e. $\delta U+\delta W=0$ \citep[e.g.][]{VAL2017}. In the regime where compressive effects are small, this then becomes $\delta U \approx 0$ and for an adiabatic process reduces to a conservation of total pressure equation. 
As we have assumed an initial constant pressure, this implies that the mixing gives
\begin{equation}
p_{\rm mixed}=p.
\end{equation}
As we see later, the latter assumption of small compressibility can be violated, but this provides a good starting point to estimate the characteristic pressure in the mixing layer though further work is necessary to make it completely accurate in all cases.
This implies that a measure of the temperature in this mixed region can be given as 
\begin{equation}
T_{\rm mixed}=\frac{p_{\rm mixed}\mu}{\rho_{\rm mixed} R_{\rm g}},
\end{equation} 
where $\mu$ is the mean molecular mass and $R_{\rm g}$ is the gas constant. Therefore:
\begin{equation}\label{mixed_temp}
T_{\rm mixed}=\sqrt{T_1 T_2}.
\end{equation}

The characteristic field strength in the mixing layer is necessarily determined by conservation of flux.
This again leads to a simple averaging to determine the field strength, which being constant in the domain just gives:
\begin{equation}\label{mixed_flux}
B_{\rm mixed}=B.
\end{equation}
%It is however important to note that total pressure calculated in the mixed region will not be the same as the external total pressure.

\subsection{Developing a simple model to estimate the kinetic energy of the mean flow}

One aim of this paper is to estimate the proportion of the energy that is extracted from the mean flow to create an upper bound on the amount of energy that can exist in turbulent flows. To do this, we must first make a model of the mean velocity field that can be used to estimate the kinetic energy that remains in the mean flow (i.e. is not available for turbulent motions) and then whatever is left over can be used as the upper limit for the turbulent kinetic energy.

\subsubsection{Mean density and velocity profiles}\label{vel_field}

The profile across the mixing layer of the averaged density and velocity would provide important information on the kinetic energy distribution in the mixing layer. 
However to develop an approximation of these profiles, further constraints and assumptions are necessary. 
The mixing layer has been placed into its zero-momentum frame, therefore any velocity profile needs to be such that this condition is maintained, on top of this conservation of mass of the layer must be observed.

To model $\langle \rho \rangle$ and $\langle v_x \rangle$ we develop an approximate polynomial solution, based on basic rules developed for the mixing layer.
We first apply the condition that both the $\langle \rho \rangle$ and $\langle \rho \rangle\langle v_x \rangle$ are continuous, i.e. at either edge of the mixing layer they take the values of the background density and flow.
This is $\rho_1$ and $\rho_2$ for the density and the results given in Equations \ref{left_vel} and \ref{right_vel} for the velocity.
Conservation of mass demands that:
\begin{equation}
\int_{-1+Y'}^{1+Y'}\langle \rho \rangle \mathrm{d}y'=2\sqrt{\rho_1\rho_2},
\end{equation}
where $\mathrm{d}y'=\mathrm{d}y/l$.
The conservation of momentum is more complex, with the true statement of the conservation of momentum in the layer gives 
\begin{equation}
\int_{-1+Y'}^{1+Y'}\langle \rho \rangle\langle v_x \rangle+\langle \rho' v_x' \rangle \mathrm{d}y' =0,
\end{equation}
where the primes denote the fluctuating component. We assume that the fluctuations in the density and velocity fields are {essentially} uncorrelated meaning that the magnitude of the fluctuating term goes to zero when integrated across the layer so our condition becomes:
\begin{equation}
\int_{-1+Y'}^{1+Y'}\langle \rho\rangle \langle v_x \rangle \mathrm{d}y'=0.
\end{equation}
We then require that the distributions of $\langle \rho \rangle$ and $\langle \rho \rangle\langle v_x \rangle$ vary monotonically.
We also prescribe that as the mixing of the momentum happens in conjunction with the mixing of the density, the $y$ position where $\langle \rho \rangle=\rho_{\rm mixed}$ is the same as the position where  $\langle \rho \rangle\langle v_x \rangle=0$.
Finally, as has been key to the derivations performed so far, we have assumed that the nonlinear dynamics works to release as much of the kinetic energy from the mean flow as possible, i.e. that 
\begin{equation}
\int_{-1+Y'}^{1+Y'}\tfrac{1}{2}\langle \rho \rangle \langle v_x \rangle^2 \mathrm{d}y'
\end{equation}
is minimised.

{After applying these rules, third-order polynomials for $\langle \rho \rangle$ and $\langle v_x \rangle$ are determined under the constraint of energy minimisation.
The steps applied to this minimisation are as follows:
\begin{itemize}
\item Using a uniform grid, the mixing layer is discretised into 1001 grid points.
\item Taking each point in the grid in turn as the point where $\langle \rho \rangle=\sqrt{\rho_1\rho_2}$, the polynomial for the density distribution is determined using the rules on the total density and the density at each end of the layer.
\item For a density distribution where the gradient is positive throughout the layer, the velocity distribution is then calculated.
\item Using the grid point where $\langle \rho \rangle=\sqrt{\rho_1\rho_2}$, we set $\langle v_x \rangle=0$ and determine the $\langle v_x \rangle$ solution based on the other constraints on the momentum listed above.
\item All grid points on the grid are cycled through, and the grid point where the constraints on the distribution are satisfied and is associated with the least energy in the mean flow is selected.
\end{itemize}
}

\begin{figure*}
  \begin{center}
\includegraphics[width=8cm]{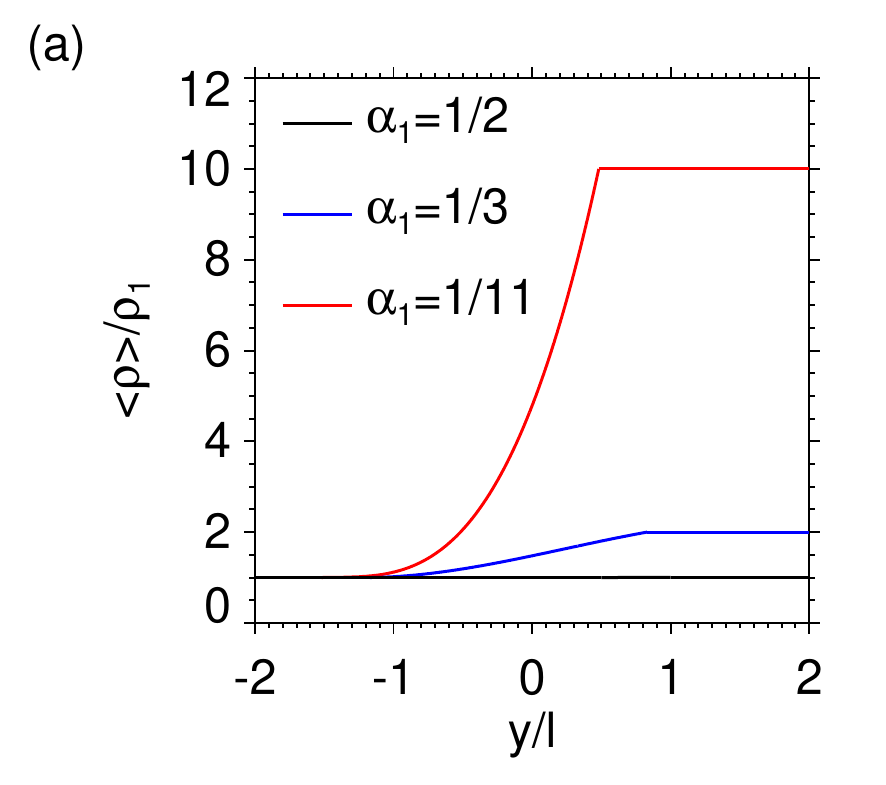}
\includegraphics[width=8cm]{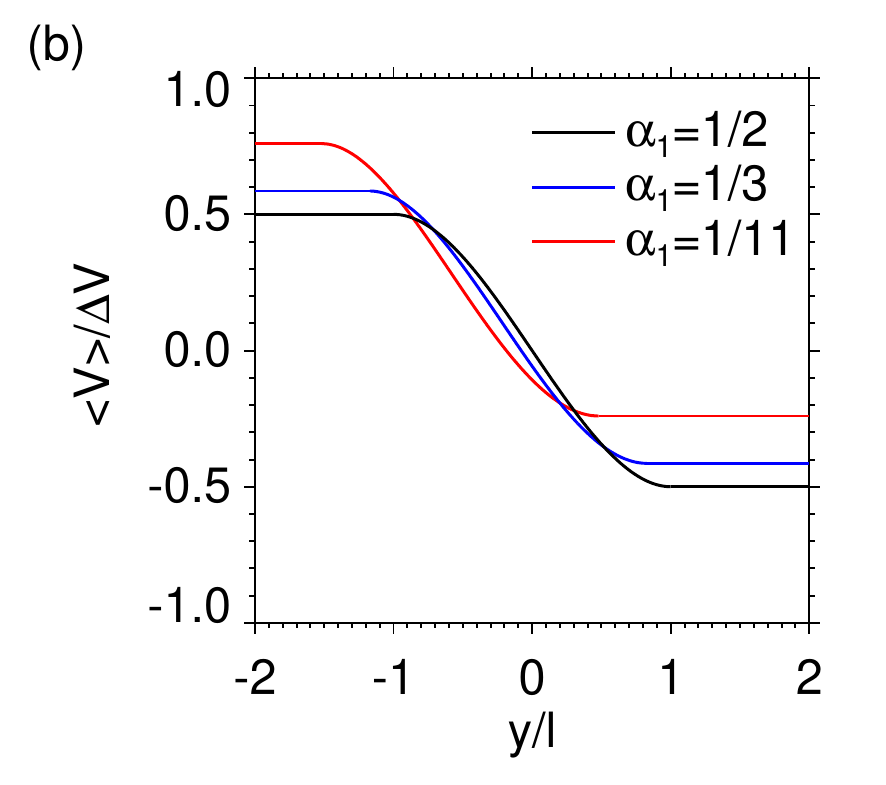}
  \end{center}
  \caption{Density and velocity field of the mixed region for a calculation of $\Delta V=1$. The red line has $\alpha_1=1/11$, the blue line has $\alpha_1=1/3$ and the black line has $\alpha=1/2$.}
\label{vel_prof}
\end{figure*}

The approximate solutions are shown in Fig. \ref{vel_prof} with the density distribution for our model using three different values of $\alpha_1$ in panel (a) and the same for the velocity in panel (b).
The shift in the position of the layer aside, there are some interesting effects from the changing of the density contrast, the most important being that the density distribution becomes heavily skewed. This results in the point where $\langle \rho \rangle=\rho_{\rm mixed}$ becoming closer and closer to the high-density edge of the mixing layer.
It is worth noting that the distribution of the $\alpha_1=1/2$ solution for $\langle v_x\rangle$ is similar to that of the error function, which is important as this is the classic solution (confirmed via comparison with experimental data) used to explain turbulence developing between two flows \citep[e.g.][]{WIN1974}.
There is one important difference: as our model is separated into three layers, if a continuous, smooth function was to be used to explain the distribution it would be non-analytic. 

\begin{figure*}
  \begin{center}
\includegraphics[width=8cm]{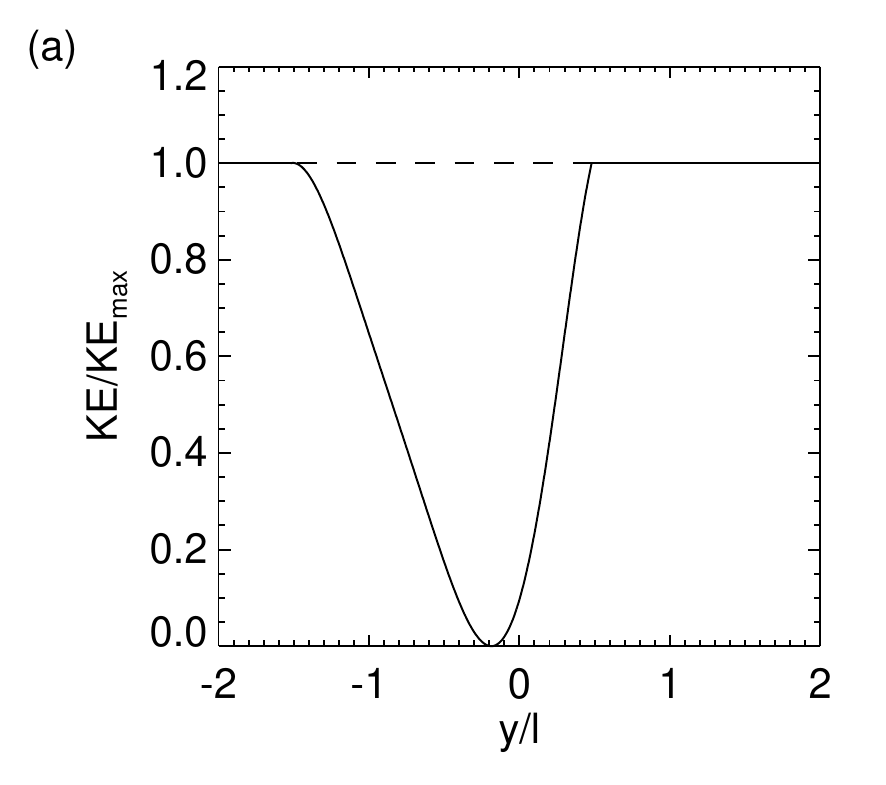}
\includegraphics[width=8cm]{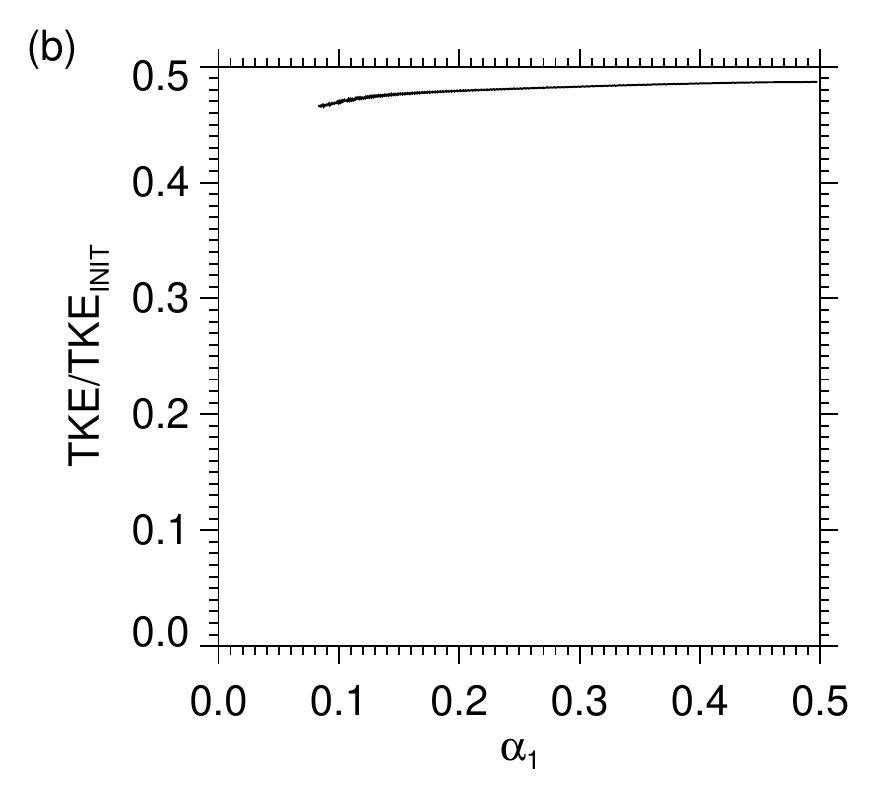}
  \end{center}
  \caption{Plot of the kinetic energy of the mean flow normalised by the value of the kinetic energy outside of the mixing layer for the case where $\alpha_1=1/11$. The dashed line gives the initial kinetic energy distribution (left). Also plotted is the ratio of the mean-flow total kinetic energy (${\rm TKE}$) to the initial total kinetic energy (${\rm TKE_{INIT}}$) against $\alpha_1$.}
\label{KE_prof}
\end{figure*}

The kinetic energy distribution in the mixing layer is given by:
\begin{equation}
{\rm KE}_{\rm mixed}(y)=\tfrac{1}{2}\langle \rho \rangle \langle v_x \rangle^2.
\end{equation}
Because of the continuity of the density and velocity, this is also a continuous distribution.
An example of the kinetic energy distribution, in this case for $\alpha_1=1/11$, is presented in panel (a) of Figure \ref{KE_prof}.
The total kinetic energy of this mean component is also relatively simple to calculate, using $\langle \rho\rangle$ and $\langle x\rangle$ and integrating over $y$ gives:
\begin{align}
{\rm TKE}=& \frac{1}{2}\int_{-1+Y'}^{1+Y'}\langle \rho \rangle \langle v_x \rangle^2 \mathrm{d}y'.
\end{align}
This can be compared to the initial kinetic energy of the band, i.e. the energy before mixing:
\begin{align}
{\rm TKE}_{\rm INIT}=&\rho_{\rm mixed}\frac{\Delta V^2 (\alpha_1\alpha_2)^{1/2}}{(\sqrt{\alpha_1}+\sqrt{\alpha_2})^2}.
\end{align}
The comparison between these two is presented for a range of $\alpha_1$ values in panel (b) of Figure \ref{KE_prof}, where it is clear that this ratio is always less than $0.5$ and becoming smaller as $\alpha_1$ tends to 0. Therefore we can approximate ${\rm TKE}$ by
\begin{equation}
{\rm TKE}\approx \frac{1}{2}\rho_{\rm mixed}\frac{\Delta V^2 (\alpha_1\alpha_2)^{1/2}}{(\sqrt{\alpha_1}+\sqrt{\alpha_2})^2}.
\end{equation}
%Note, as when looking at the kinetic energy before, the choice of reference frame becomes important when looking at the second order components and so it is important to work in the rest frame of the mixing layer.
{We note that the range of $\alpha_1$ used in panel (b) of Figure \ref{KE_prof} is reduced because the accuracy of the third order polynomial approximation deteriorates at small $\alpha_1$ values due to the low order of the polynomial used. This implies that further constraints exist on the distributions, likely to be related to the various orders of the derivatives of the density, momentum and kinetic energy at the edges of the layer, that could be used to extend this approximation to higher order and with that lower $\alpha_1$.}

\subsection{Estimate of the fluctuating energy component}\label{diss}

The next question we would like to approach is: how much energy is there available to dissipate via the turbulent creation of small scales?
In other words, what energy is there available for heating from the fluctuations of the velocity field, and in an MHD system the magnetic field, around their average values.
The density, and with it the temperature, of the mixed layer does not depend of the amount of free kinetic or magnetic energy there is to dissipate, it is purely based on the density and temperatures of the regions before they are mixed.
However, the total dissipation that can occur is highly dependent on these.

In Section \ref{vel_field} we formulated the mean velocity field of the mixed region.
This velocity field is related to the average velocity profile and does not include the fluctuating component of the velocity field, often referred to as the turbulent component.
The profile shows the lowest energy state the velocity field can reach without further thickening of the mixing layer through development of larger vorticies or through viscous dissipation, i.e. it is related to how much energy the instability can release through turbulent motions.
Therefore, the energy held in the fluctuations can be estimated by the initial energy available for the instability minus the newly developed mean kinetic energy profile.
The total kinetic energy of the fluctuating component of the velocity field is given as:
\begin{equation}
{\rm TKE}_{\rm turb}\sim\frac{1}{2}\rho_{\rm mixed}\frac{\Delta V^2 (\alpha_1\alpha_2)^{1/2}}{(\sqrt{\alpha_1}+\sqrt{\alpha_2})^2},
\end{equation}
which is minimised for large density differences.
The average turbulent kinetic energy across the layer is given as: 
\begin{equation}
\overline{\rm KE}_{\rm turb}\sim\frac{1}{4}\rho_{\rm mixed}\frac{\Delta V^2 (\alpha_1\alpha_2)^{1/2}}{(\sqrt{\alpha_1}+\sqrt{\alpha_2})^2}.
\end{equation}
The characteristic magnitude of the velocity fluctuations is given by:
\begin{equation}\label{turb_rms_vel}
V_{\rm turb, RMS}\sim \sqrt{\frac{2}{\rho_{\rm mixed}}\overline{{\rm KE}}_{\rm turb}}\sim\sqrt{\frac{1}{2}}\frac{\Delta V (\alpha_1\alpha_2)^{1/4}}{(\sqrt{\alpha_1}+\sqrt{\alpha_2})}.
\end{equation}

The average increase in internal energy of an ideal gas as a result of processes other than mixing is given by $\overline{\Delta p}/(\gamma-1)$, i.e. the heating must come from the dissipation of the turbulent kinetic energy, therefore from energy conservation we know:
\begin{align}
\frac{\overline{\Delta T}}{T_{\rm mixed}}\sim
\frac{\overline{\Delta p}}{p_{\rm mixed}}\approx &\frac{1}{4}(\gamma-1)\frac{(\alpha_1\alpha_2)^{1/2}\Delta V^2}{(\sqrt{\alpha_1}+\sqrt{\alpha_2})^2}\frac{\rho_{\rm mixed}}{p_{\rm mixed}}\nonumber\\
=&\frac{\gamma}{4}(\gamma-1)\frac{(\alpha_1\alpha_2)^{1/2}}{(\sqrt{\alpha_1}+\sqrt{\alpha_2})^2}\frac{\Delta V^2}{C_{\rm S, mix}^2}\nonumber\\
\sim&\frac{1}{4}\frac{(\alpha_1\alpha_2)^{1/2}}{(\sqrt{\alpha_1}+\sqrt{\alpha_2})^2}\frac{\Delta V^2}{C_{\rm S, mix}^2}\nonumber\\
\le&\frac{1}{16}\frac{\Delta V^2}{C_{\rm S, mix}^2}=\frac{1}{16}M^2\label{heating}
\end{align}
where $\gamma$ is the adiabatic index, $C_{\rm S, mix}$ is the characteristic sound speed of the mixed region where the energy is being dissipated and $M$ is the Mach number.
Therefore, if we know the velocity shear and can estimate the sound speed of the mixed region then we can give a bound for the average heating. If the density ratio is also relatively well constrained then this can be used to accurately estimate the heating.

\subsubsection{Estimating the effect of the turbulent pressures}\label{compress}

Both the dynamic and magnetic pressures that result from the fluctuating velocity and magnetic field, respectively, in the mixing layer can work to expand the mixing layer by acting to add to the total pressure of that region.
The magnitude of these two turbulent pressures is intrinsically related to the amount of kinetic energy that is taken from the large scale shear flow and put into the turbulent components of the velocity and magnetic field.
The characteristic Mach number of the instability or the characteristic Alfv\'{e}n Mach number of the instability depending on whether the system is high or low $\beta$, where if these characteristic numbers are small then it can be expected that this effect is negligible.
See the results in Section \ref{diss} for estimates of the kinetic energy that can be transferred to these fluctuations (and with it an approximation of the total turbulent pressure they can create).
As the turbulent pressures are representative of the energy available for dissipation, similar arguments can be formed for loss of force balance through the pressure increase through dissipative heating.

%{\rm needs developing?}

\subsection{Timescales for mixing}

The previous parts of this section focussed on the quantities of the mixed layer. 
These are the quantities that the mixing is driving the system to achieve and represent the {{mean values of} the layer during the mixing process.
However, there remains a very important question: What is the timescale over which the mixing occurs?

To answer this, it is necessary to have a measure of the magnitude of the velocity fluctuations in the mixing layer, which can be taken from the fluctuating energy component.
Using the estimate for the turbulent velocity RMS given in Equation \ref{turb_rms_vel}, the mixing time can be approximated by an eddy turnover time, i.e.:
\begin{equation}\label{mix_time}
\tau_{\rm mixing}\approx\frac{2l}{V_{\rm turb, RMS}}\ge\frac{2l}{\Delta V }\sqrt{2}\frac{\sqrt{\alpha_1}+\sqrt{\alpha_2}}{(\alpha_1\alpha_2)^{1/4}},
\end{equation}
which gives a measure of how long it takes to mix the region.

This mixing time will strongly correlate to the dissipation time scale in the limit where $\tau_{\rm mixing} \ll \tau_{\rm viscous}$.
In a turbulent system, which is likely to form under the previously stated conditions, the nature of the cascade implies that the longest time scales are those at the largest scale.
Therefore, as with the mixing the dissipation rate is connected to, albeit in a complex fashion, to the time scales at the largest scale.
As such the mixing time can also be used as a very approximate measure of the lower limit of the dissipation time scale in turbulent mixing.

The addition of a magnetic field to the problem adds a number of other considerations.
In high Lundquist number flows the fluid is strongly tied to the magnetic field, which inhibits mixing.
Therefore, as discussed in the introduction, to have quick efficient mixing, magnetic reconnection leading to the disruption of {vortices} becomes necessary.
In the 3D simulations of \citet{ANT2015}, many current sheets were found to form (a necessary condition for magnetic reconnection) showing that mixing is possible in the regime we are interested. So assuming that the current sheets form on an eddy-turnover timescale, i.e. the timescale we have estimated, then this will still hold as an approximate mixing timescale.

\section{Numerical simulations of Kelvin--Helmholtz Mixing}

To both confirm the key predictions and evaluate the limits of this model, we present the results from a 3D MHD simulation of Kelvin--Helmholtz mixing.

\subsection{Setup}

Using the (P\underline{I}P) code \citep{HILL2016} we solve the non-dimensionalised ideal MHD equations:
\begin{align}
\frac{\partial\rho}{\partial t}&+\nabla\cdot(\rho\mathbf{v})=0, \\
\frac{\partial}{\partial t}(\rho\mathbf{v})+\nabla\cdot&\left(\rho\mathbf{v}\mathbf{v}+P\mathbf{I}-{\mathbf{BB}}+\frac{\mathbf{B}^2}{2}\mathbf{I}\right)=0,\\
\frac{\partial}{\partial t}\left( e+\frac{B^2}{2} \right)&+ \nabla\cdot\left[\mathbf{v}(e+P)-(\mathbf{v}\times\mathbf{B})\times\mathbf{B}\right]=0, \\
\frac{\partial \mathbf{B}}{\partial t}&-\nabla \times (\mathbf{v}\times \mathbf{B})=0,\label{ind_eqn}\\
\nabla\cdot\mathbf{B}&= 0,\\
e & \equiv \frac{P}{\gamma-1}+\frac{1}{2}\rho v^2.
\end{align}
This system of equations has been non-dimensionalised in the following way: The velocity $\mathbf{v}$ has been non-dimensionalised using the sound speed $C_{\rm s}$, the density $\rho$ by a reference density $\rho_0$, and the lengthscale by an arbitrary length $L$. Therefore, time $t$ is nondimensionalised by $L/C_{\rm s}=\tau_{\rm D}$, the pressure $P$ by $C_{\rm s}^2\rho_0$, and the magnetic field $\mathbf{B}$ by $B_0/\sqrt{4\pi}=C_{\rm s}\sqrt{\rho_0}$. Here $\gamma$ is the adiabatic index and $\beta$ is plasma $\beta$ (the ratio of gas to magnetic pressure calculated using the total gas pressure of the fluids).
We assume the ideal gas law which in non-dimensional form becomes $T=\frac{P\gamma}{\rho}$.

The scheme used is a fourth-order central difference scheme using a four-step Runge-Kutta scheme for the time integration.
For stability of the scheme we employ the artificial viscosity/diffusion as described in \citet{REM2009}. As this is a conservative scheme, the artificial dissipation results in an internal energy increase matching the amount of energy that has been dissipated.

The initial conditions used for the simulation present match as closely as possible to both the model developed and include the general characteristics in terms of speed and density of the flows believed to exist in the observed prominence threads.
Initial conditions for MHD simulations are given by:
\begin{align}
V_{x}(y)=&V_1+\frac{\Delta V}{2}\left(\tanh\left(\frac{y}{0.003} \right)+1\right),\\
\rho(y)=&\rho_1+\frac{\rho_2-\rho_1}{2}\left(\tanh\left(\frac{y}{0.003} \right)+1\right),\\
p(y)=&p_1=\frac{1}{\gamma},\\
B_z(y)=&B_1=\sqrt{\frac{2}{\gamma \beta}},
\end{align}
where $\rho_1=1$, $\rho_2=10$, $p_1=1/\gamma$, $V_1=10/11\times\sqrt{0.1}$ and $\Delta V=\sqrt{0.1}$ (i.e. the sound speed of the cool region) which initiates the instability in its approximately its linear reference frame (which is different from the frame we predict the nonlinear dynamics will be at rest). We set the plasma $\beta$ to be $\beta=2 p_1/B_1^2=0.3$ and take $\gamma=5/3$.
The instability is seeded with a random noise perturbation in $v_y$ at the level of $0.01\Delta V$.

The simulation is solved in the spatial domain of $x=[-0.4,0.4]$, $y=[-1.5,0.5]$ and $z=[-8,8]$ using $160\times 400\times 800$ grid points.
{Here we have taken the length of the $z$ direction to be much greater than that of either the $x$ or $y$ directions.
This is chosen because without a sufficient length along the magnetic field, the vortices will not be able to sufficiently wrap up the magnetic field and will not disrupt as there are insufficient currents for magnetic reconnection to take place.
The lengthscale required for disruption to be possible can be estimated by requiring that the rotation rate of the vortex is greater than the frequency of an Alfv\'{e}n wave, i.e.:
\begin{equation}
\frac{V_{\rm turb,RMS}}{2l}>\frac{V_{\rm A}}{L},
\end{equation}
where $2l$ is the width of the vortex, $L$ is the length along the magnetic field and $V_{\rm A}=B_1/\sqrt{\rho_{\rm mixed}}$.
Therefore,
\begin{equation}\label{length_wrap}
L>\frac{V_{\rm A}2l}{V_{\rm turb,RMS}}=\frac{B_1}{\sqrt{2\rho_{\rm av}}}\frac{2l}{\Delta V}{2l}\frac{\sqrt{2}(\sqrt{\alpha_1}+\sqrt{\alpha_2})}{\sqrt{\alpha_1\alpha_2}}\approx 11.8,
\end{equation}
where we have assumed a mixing layer width of $2l=1$.
As the length of our $z$ direction is greater than $11.8$ we expect the vortices to become disrupted.}

In this calculation we use the following boundary conditions.
The $x$ and $z$ boundaries are set as a periodic boundary, with a symmetric boundary that cannot be penetrated by the magnetic field for the $y$ boundary.

\subsection{Simulation results}

Figure \ref{3D_contour} shows the contour plots of the temperature distribution in the $xy$-plane at $z=0$.
These plots show the temperature structure at four different times ( $t= 0,\; 20,\; 40$ and $60$) covering the initial conditions through the early nonlinear stages toward a layer that is becoming well-mixed {(note that the lack of coherent vortex structures here is a sign that they have become disrupted)}. Here we can see that in the region $y>0$ there is an increase in temperature, and we will determine whether this is created by mixing or heating.  

\begin{figure*}
  \begin{center}
\includegraphics[height=18cm]{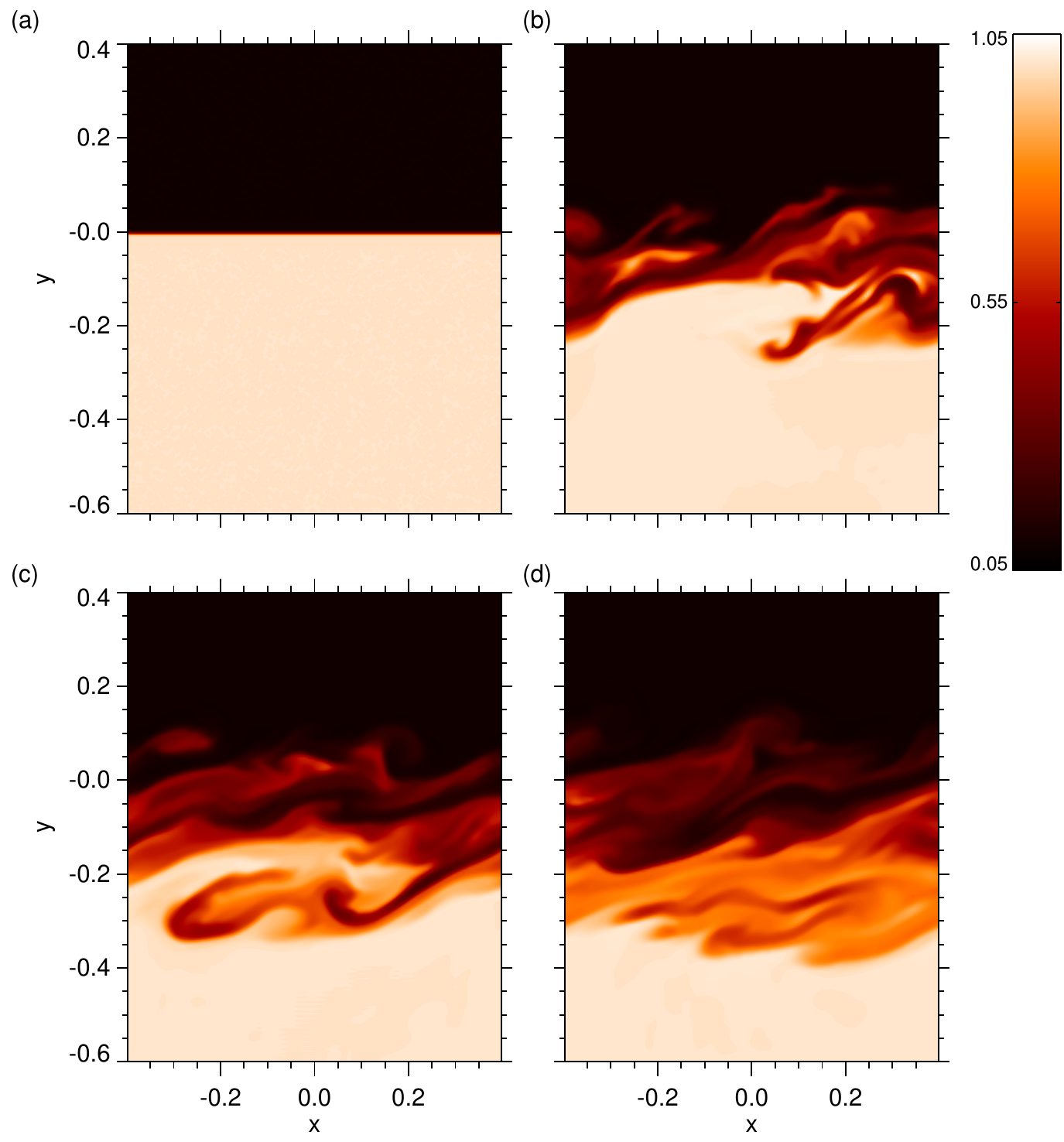}
  \end{center}
  \caption{Contour plots of the temperature distribution in the $x$-$y$ plane taken at $z=0$ for $t= 0,\; 20,\; 40$ and $60$ respectively.}
\label{3D_contour}
\end{figure*}

\begin{figure*}
  \begin{center}
\includegraphics[height=5.4cm, trim={0.4cm 0.2cm 0.4cm 0.1cm}, clip]{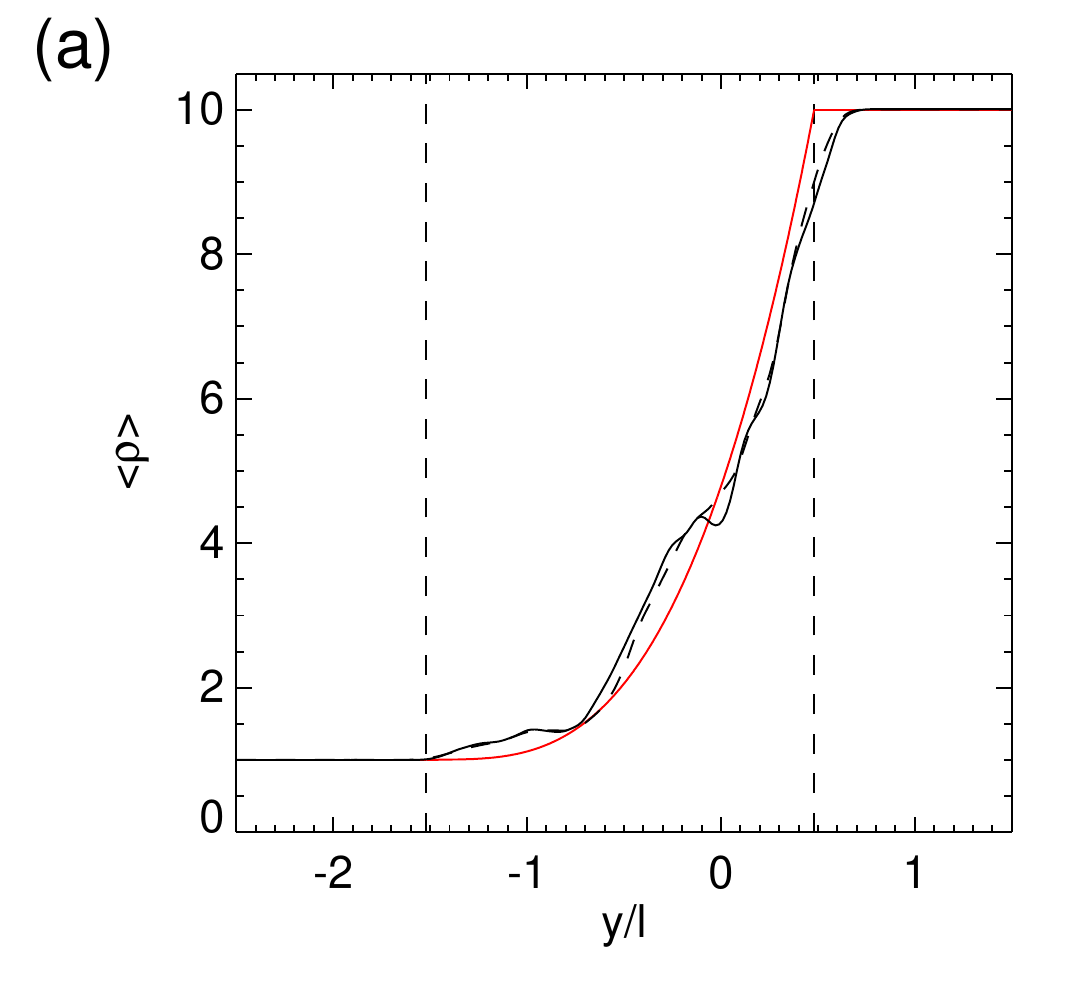}
\includegraphics[height=5.4cm, trim={0.4cm 0.2cm 0.4cm 0.1cm}, clip]{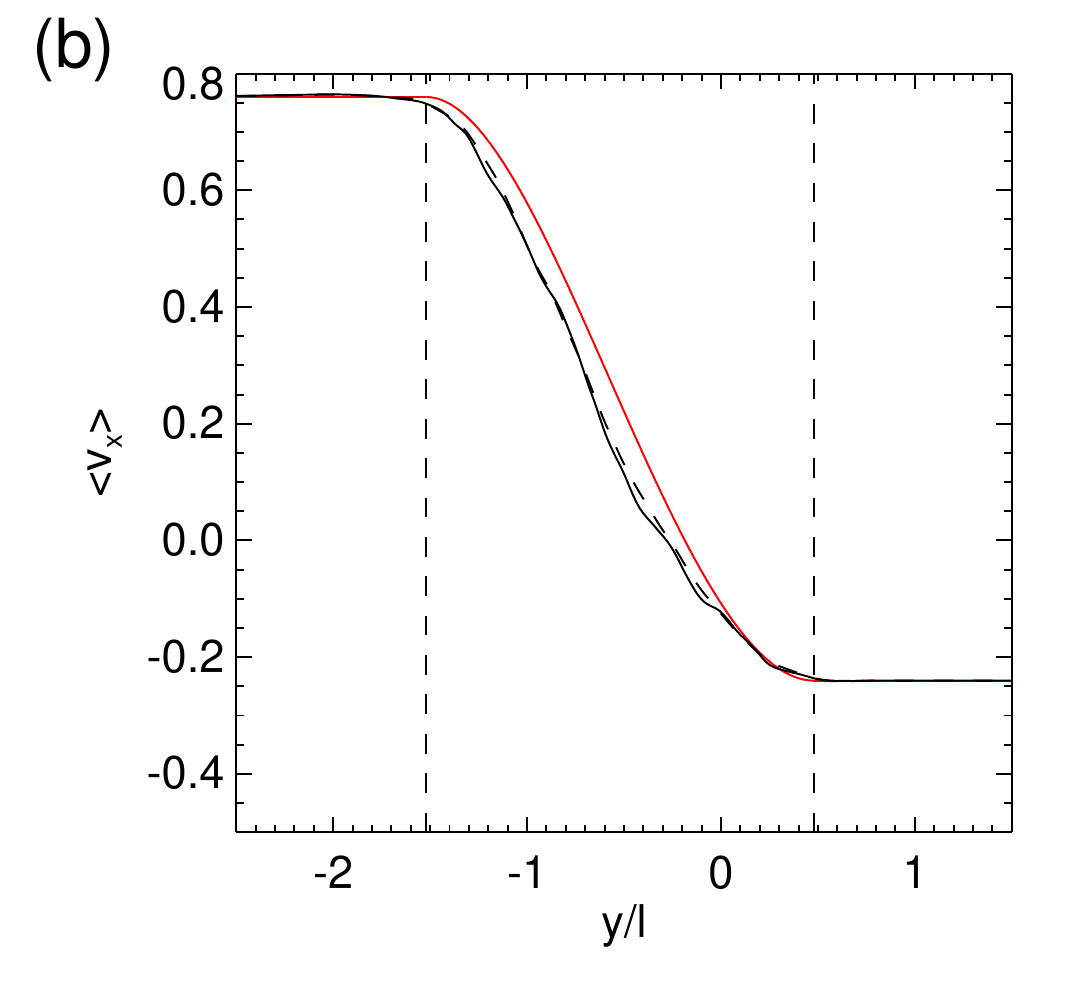}
\includegraphics[height=5.4cm, trim={0.4cm 0.2cm 0.4cm 0.1cm}, clip]{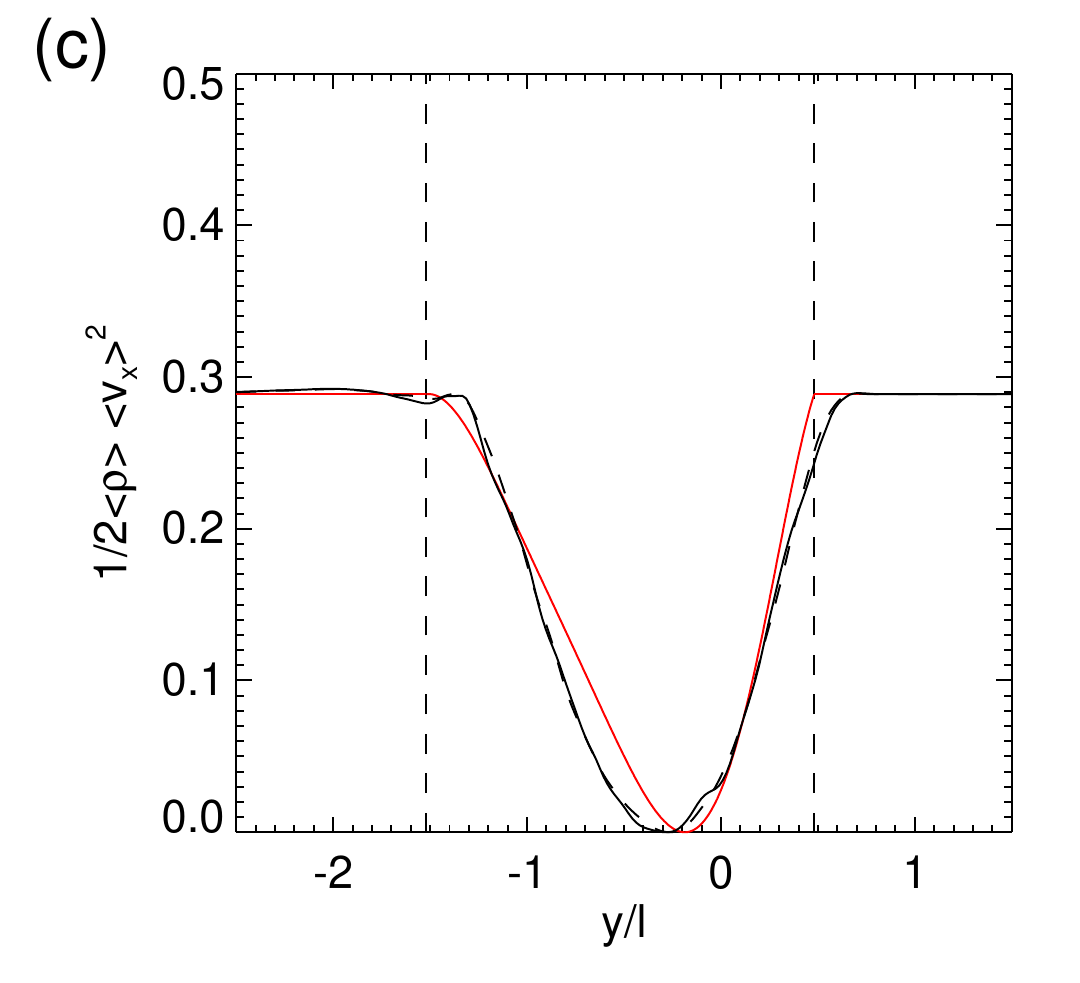}
  \end{center}
  \caption{Plots of the mean density $\langle\rho \rangle$ (a), mean velocity $\langle v_{\rm x} \rangle$ (b), and mean kinetic energy $1/2\langle \rho \rangle\langle v_{\rm x} \rangle^2$. The solid black curves give the simulation results at $t=60${, the dashed black curves give the output averaged at $4\tau_{\rm D}$ intervals between $t=40$ and $t=60$,} and the solid red lines show the model. The vertical dashed lines give the range of the model mixing layer.}
\label{mean_comps}
\end{figure*}

Figure \ref{mean_comps} shows the $x$-$z$ averaged (a) density ($\langle\rho\rangle$), (b) $x$ velocity ($\langle v_x\rangle$), and (c) mean flow kinetic energy ($\langle\rho\rangle\langle v_x\rangle^2/2$) profiles across the mixing region at $t=60$ {(solid black line)}.
Before looking at the distributions and the characteristic values of these quantities in the mixing layer, we need to state how we determined the position of the mixing layer.
For this, we look at the distribution of $\langle \rho\rangle$ and determine the point where the density departs from the minimum level by 1 per cent.
This gives the position of the $y<0$ end of the mixing layer.
As the shift in the layer is purely a function of $\alpha_1$ and $\alpha_2$, this {is determined} by the initial conditions to be (in normalised units) approximately $-0.52$.
Together these uniquely determine the position of the mixing layer associated with the minimum mean velocity.
Visually it seems that the predicted shift is a good representation of the shift in the mixing layer from the $y=0$ position.

Panel (a) of Figure \ref{mean_comps} gives the profile of $\langle \rho\rangle$ against $y$. 
The red lines show the simplified model used, which is not a perfect representation but provides a very good estimate of the $\langle \rho \rangle$ distribution. Calculating from the simulation the mean density across the whole mixing layer gives a value of $\overline{\rho}=3.21$. The predicted density from the model for the simulation parameters is $\rho_{\rm mixed}=\sqrt{10}\sim3.16$, which is a difference of less than two per cent. This small difference can be understood by the slight extension of the mixing layer on the right hand side resulting in slightly more mass existing in the layer than predicted.
Therefore, both the total mass and the spatial distribution of $\langle \rho \rangle$ are well represented.

Panel (b) of Figure \ref{mean_comps} gives the distribution of $\langle v_{\rm x}\rangle$ against $y$. 
The model for this quantity is shown in red and this provides a reasonably accurate model of the velocity profile.
We can also see that the bounds on the mixing layer for this quantity are accurate.
%The dash-triple dot lines in panels (a) and (b) show the position where $\langle \rho\rangle=\rho_{\rm mixed}$ {\bf add lines or at least develop arguments}. This position approximately corresponds to the point of zero mean velocity, highlighting why this connection was important for producing the initial model.

Looking at panel (c) of Figure \ref{mean_comps}, which shows the kinetic energy of the mean flow, in the centre of the mixing layer this energy goes to zero. The energy that has been removed from the mean flow is $0.58$ of the initial energy available in the mixing region. The red line in the panel shows the model prediction for the kinetic energy of the mean flow. The model predicts that $0.53$ of the total energy that exists in the mean flow initially in the mixing layer has disappeared from the kinetic energy of the mean flow. The distribution is narrower than the simulation results, leading to the small under estimation. However, this confirms that our model can provide a sufficiently accurate representation of the energy available for turbulent motions, and with this heating, in the nonlinear stages of the instability. {The dashed black line in the panels of this figure show the profiles achieved when time averaging as well as spatial averaging is considered. Though this appears to marginally improve the, already good, match between the model and the simulation results, the fact that the match is still not perfect highlights that there are still constraints that exist that are not yet considered in this model.}

\begin{figure}
  \begin{center}
\includegraphics[height=8cm]{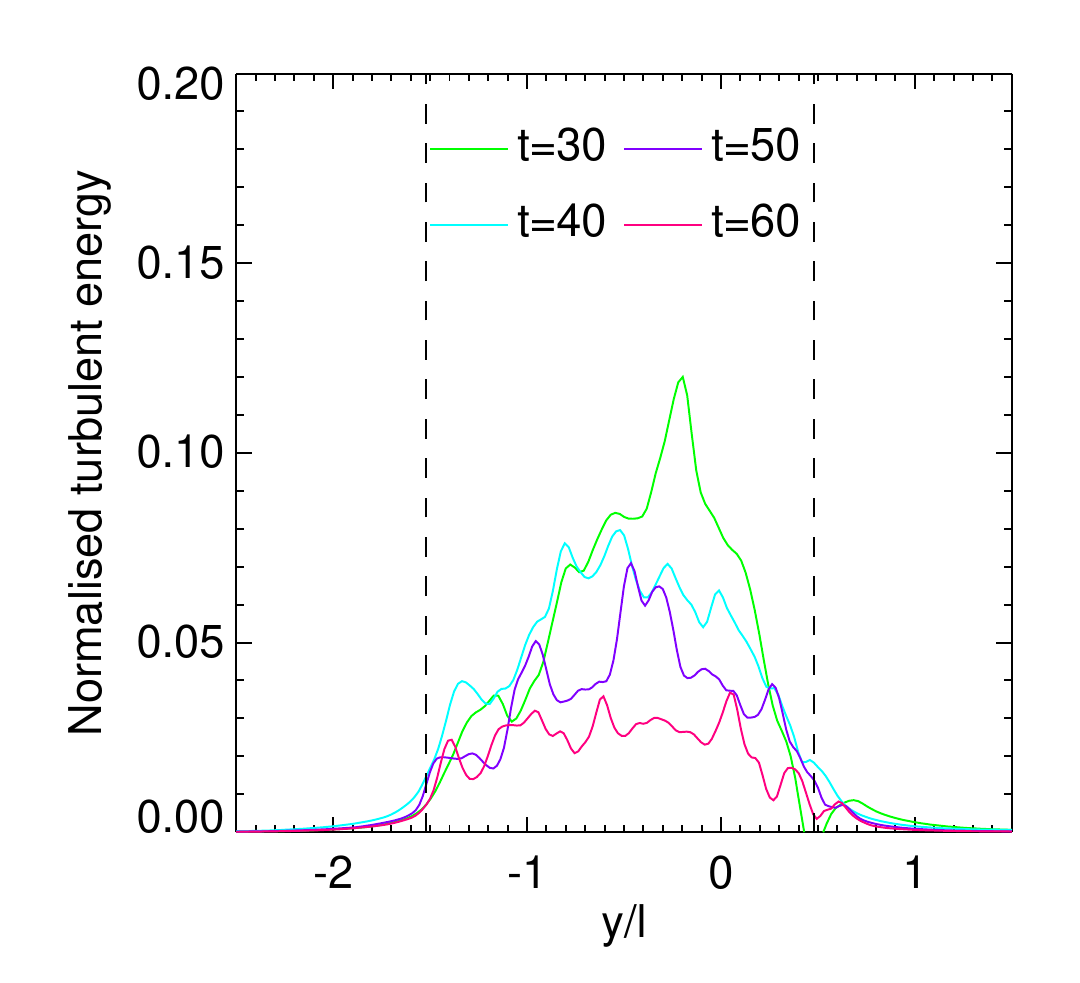}
  \end{center}
  \caption{$y$ distribution of the turbulent energy held in both the velocity and magnetic field fluctuations (averaged in both the $x$ and $z$ directions). This is plotted for a number of separate times, showing that the energy in these fluctuations is decreasing, i.e. that energy is being dissipated. Vertical dashed lines give the extent of the mixing region in the model.}
\label{turb_en}
\end{figure}

Figure \ref{turb_en} shows the distribution of the turbulent energy (i.e. the energy held in the velocity and magnetic field fluctuations) normalised by the initial kinetic energy density in the zero-momentum frame of the mixing layer at four separate times during the simulation ($t=30$, $40$, $50$ \& $60$). The trend over time is that the magnitude of the turbulent energy decreases. This is a clear signature of the dissipation of the turbulent energy.

\begin{figure}
  \begin{center}
\includegraphics[height=8cm]{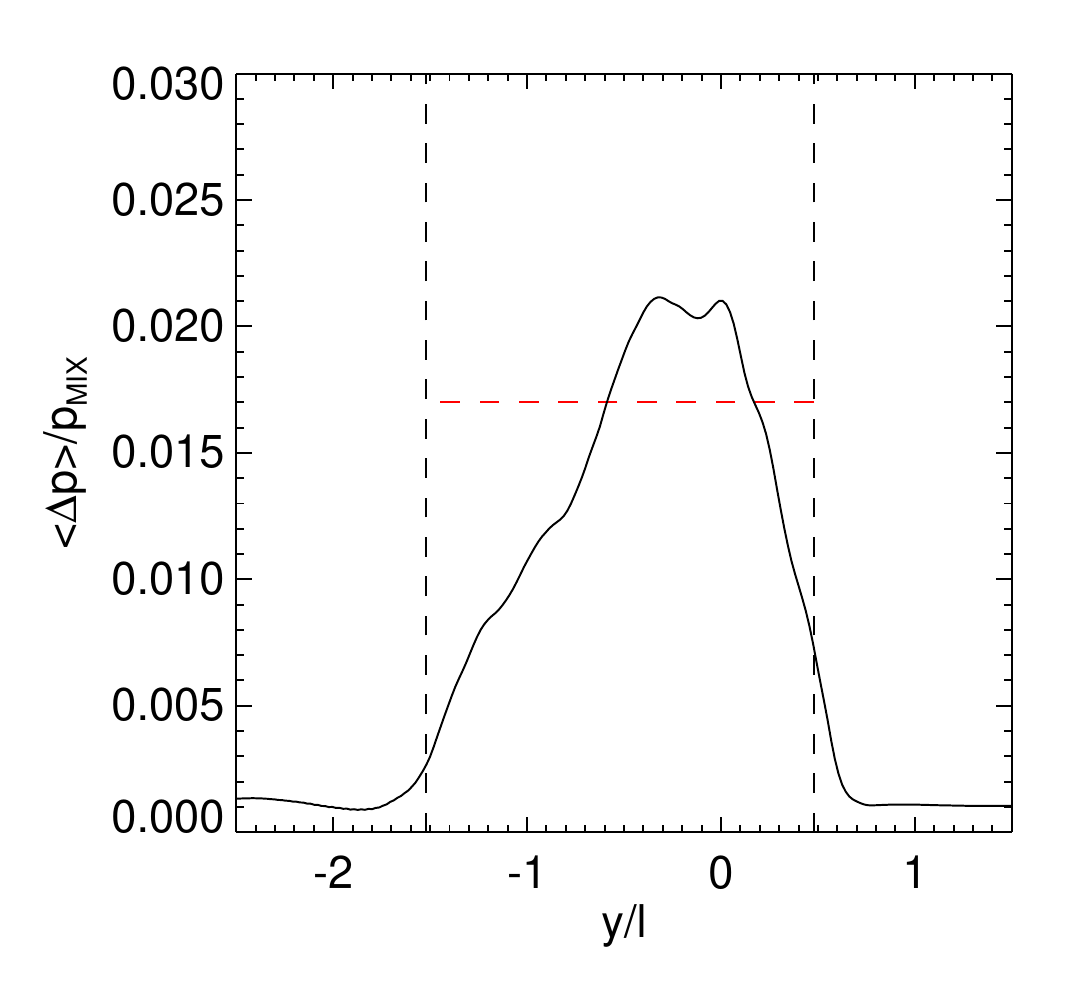}
  \end{center}
  \caption{Magnitude of the pressure fluctuations, averaged in the $x$ and $z$ directions, normalised by $p_{\rm mixed}$ at $t=60$. Vertical dashed lines give the extent of the mixing region in the model. The horizontal dashed red line gives the predicted increase in pressure as a result of turbulent heating.}
\label{mean_pres}
\end{figure}

Figure \ref{mean_pres} shows the magnitude of the pressure fluctuations $\langle\Delta p\rangle$ from $p_{\rm mixed}$ (normalised by $p_{\rm mixed}$, which in the case of this simulation $=1/\gamma=0.6$) taken at $t=60$. As can be seen the increase in the pressure that results from dissipation and any compressible effects peaks at approximately two per cent of $p_{\rm mixed}$. The horizontal dashed red line gives the value of the average pressure increase from heating from the model as given in Equation \ref{heating}, which is calculated as $\sim 0.017$. Visibly this line can been seen as a fair estimate of the average increase of the pressure through heating. The actual value of $\overline{\Delta p}/p_{\rm mixed}$ from the simulation is $\overline{\Delta p}/p_{\rm mixed}=0.014$. 
This is less than the prediction, but as there is still turbulent energy that can be dissipated (see Figure \ref{turb_en}) then this could rise further over time.
Ultimately, due to the marginally increased extraction of mean energy in the simulation we would expect that it would reach a value slightly above that of the model.
Nonetheless, the prediction is a fair reflection of the increase in internal energy in the mixing layer.
The clear conclusion from this is that the temperature of the mixing layer in the simulation is determined by the mixing process and not by any heating in the simulation {(see also Section \ref{sim_comp})}.

\section{Application to Prominence observations - Mixing VS Heating}

To apply the results of the model presented in Section \ref{phem_model} to the observations of \citet{OKA2015}, we need to use some characteristic values for the temperature, density and velocity.
We use $10^{-15}$\,g\,cm$^{-3}$ and $10^6$\,K for the coronal density and temperature, and $10^{-13}$\,g\,cm$^{-3}$ and $10^4$\,K for the density and temperature of the prominence thread, which means that we take a constant pressure between the two regions.
Also, we consider shear flows of magnitude $10$\,km\,s$^{-1}$.

The first step is to assess the thermodynamic properties expected of a mixing layer between prominence material and the corona.
For a Kelvin--Helmholtz mixing layer between these two, the characteristic density of this layer would be $\rho_{\rm mixed}=\sqrt{10^{-13}\times10^{-15}}$\,g\,cm$^{-3}=10^{-14}$\,g\,cm$^{-3}$.
The temperature under these conditions is also determined by the geometric mean so $T_{\rm mixed}=\sqrt{10^4\times10^6}$\,K$=10^5$\,K.
When this process is observed in the cool ($\sim 10^4$\,K) and warm ($\sim 10^5$\,K) passbands of IRIS, this mixing would result in cool material being removed by the mixing resulting in high intensity material disappearing from that passband. As the warmer material forms in the mixing region intensity would increase in warm passbands.
{This could be expected to observationally lead to the thinning of the prominence thread in cool lines, with the thickening of the transition region of the prominence thread when observed in warm lines.}

{It is worth noting that to achieve the mixing temperatures predicted by this model, locally the plasma has to relax from the two different particle distributions that make up the prominence and coronal plasmas to a single distribution that is at transition region densities and temperatures. In MHD simulations, a single Maxwellian distribution in each pixel is assumed to form instantaneously, but for the prominence corona system it takes sufficient particle collisions between the cooler and hotter particles to relax to a single temperature particle distribution in a local area.}

{Along magnetic field lines in the solar corona, thermal conduction can effectively perform this task, but across field lines the conduction is significantly reduced. To make the mixing temperatures, heat transport across the magnetic field is essential. To make the mixing across the field more efficient, it is necessary to break the connectivity of the magnetic field to allow thermal conduction to transfer heat from the hotter to the cooler components of the mixing layer. As long as the length of the fieldlines hosting the prominence thread are sufficiently longer than the width of the mixing layer (see Equation \ref{length_wrap} for an estimate), then the instability can wrap up the magnetic fields to produce the reconnection required to allow field-aligned thermal conduction to become important. For a $100$\,km width mixing layer, this length would be $L\approx 4\times 10^3$\,km, which is sufficiently small to allow these dynamics to occur on a prominence thread. Due to the formation of many secondary vortices as a result of the instability as it nonlinearly develops meaning smaller scales both along and across the field, the dynamic evolution of the vortices naturally produce many small scale current sheets \citep[e.g.][]{ANT2015}, which mean that reconnection as a result of the vortex evolution can be important for the thermalisation of the plasma.}

{An alternative method by which turbulent heat transport could occur is as a result of the drift of neutral atoms across the magnetic field. In the dense, cool material of prominences, the degree of ionisation has not been exactly determined but both ranges of the ionisation fraction of 0.2 to 0.9 \citep{ENG1990,LAB2010}, or ratios of electron to neutral hydrogen density in the range 0.1 to 10 \citep{PATS2002} have been reported. The work of \citet{HILL2019b} highlights the role of the motion of neutral particles across the magnetic field for heat transport. As the neutral particles drift across the magnetic field, through collisional coupling with the local plasma they meet they can act as a heat sink for the hotter material they interact with resulting in a transfer heat. Because the relaxation to a single temperature distribution due to thermal conduction is more effective in the hot, low-density coronal plasma, and the relaxation by ion-neutral drift is more important for regions of the mixing layer that have more cool material, it is likely that both of these mechanisms could be important in this mixing process in the solar corona.}

We can use Equation \ref{heating} to estimate the heating by the Kelvin--Helmholtz instability of prominence material in the solar corona.
For this we can use a velocity shear of $10$\,km\,s$^{-1}$ and for a temperature of $10^5$\,K the sound speed is $33$\,km\,s$^{-1}$. 
Combining these with the appropriate densities this would give an increase in temperature of the fluid of $\Delta T < 0.003 T_{\rm mixed}$.
That is to say, the energy available for rapid heating via turbulent dissipation can possibly result in a temperature increase that is only a fraction of a per cent of the temperature achieved through mixing, i.e. the possible heating is not significant compared to the mixing in this situation because there is just not enough energy available for dissipation. 

Using the estimate for the time scale given in Equation \ref{mix_time}, we can estimate the observable time scales for these processes.
Taking a mixing region of half-width $100$\,km would result in a lower estimate for both the mixing time scale and the heating time scale of $\sim100$\,s (note this increases linearly with increases in the half-width).
Based on the dominance of mixing, it would be sensible to assume that this process would take a few hundreds of seconds to significantly reduce the cool intensity of a prominence thread whilst producing the warmer material at a similar rate. 
Combining this time scale with the total energy available for dissipation gives an energy dissipation rate of $\sim10^{-4}$\,erg\,cm$^{-3}$\,s$^{-1}$.
Note that this is much larger than the $10^{-8}$\,erg\,cm$^{-3}$\,s$^{-1}$ estimated for quiescent prominence turbulence by \citet{HILL2017}, though this difference is mostly due to the localisation assumed for the turbulence in this study and the larger velocities used.

\subsection{Radiative losses and the possibility of coronal cooling}\label{losses}

One important consequence of the dominant role of mixing, is that, while not efficiently heating the system, it alters the temperature of the plasma changing the radiative losses.
{As shown in Figure 2 of \citet{ANZ2008} for example, the loss function $\Lambda(T)$ for optically thin radiative losses for coronal plasma at constant pressure vary with temperature, with the total radiative losses given by $R=n^2\Lambda(T)$ (where $n$ is the number density).}
For the mixing of cool ($10^4$\,K) prominence plasma with hot ($10^6$\,K) coronal plasma, the characteristic temperature of the mixing layer is $10^5$\,K.
Their calculations show that the radiative losses from the mixing region are going to be greater than the losses for either of the plasmas before mixing.

For the coronal plasma we are using, the time scale for radiative loss is approximately $10^3$\,s, and for the prominence plasma this becomes approximately $1$\,s (though it should be noted that the use of optically thin radiation to model the optically thick radiative losses of the prominence strongly underestimates this time scale).
The result of the mixing gives a time scale of approximately $1$\,s for what will be an optically thin plasma, which equates to an energy loss rate of $\sim 0.1$\,erg\,cm$^{-3}$\,s$^{-1}$.
Therefore, the cooling times have been drastically reduced by mixing.
When compared to the heating rate as a result of turbulence, it is clear that the cooling rate dominates this, meaning that even though there is heating occurring, thermal energy is being lost from the system as a result of the occurrence of the Kelvin--Helmholtz instability faster than it is being replaced.
Therefore, the overall result of the mixing process is more likely to be cooling of the solar corona than heating.

{It is important to note that the estimates for the change in the radiative losses presented above were calculated using $\rho_{\rm mixed}$ and $T_{\rm mixed}$. However, these are just characteristic values for the mixing layer, and as shown in Figure \ref{vel_prof} for large density contrasts the density, and with it the temperature, has a nonlinear distribution across the mixing layer. Therefore these estimates should only be taken as characteristic values to highlight how the cooling timescale of the prominence corona system will evolve as a result of the KHi.}

{This estimate of the cooling time is based on the fluids having become well-mixed, but even before this process has taken place, i.e. in the early stages of the instability when vortices are forming but the fluids remain relatively distinct, it is likely that the radiative losses of the prominence-corona system would increase. The high temperature component of this system is optically thin, so the corrugation of the boundary between the two fluids does not change the losses from this material. However, the optically thick emission from the prominence material is determined in part by the surface area through which the photons can escape \citep[see the shell emission found for optically thick lines from radiative transfer models of simulations presented in Figure 8 of][]{OKA2015}. }

{Taking the linear instability to have reached a displacement of the boundary of $1/k$, i.e. the instability will be developing nonlinearities \citep[e.g.][]{HILL2019b}, and using the plane wave solution of the linear instability, so we can assume a sine wave form of the boundary displacement, the surface area of the boundary increases by approximately 25\%. Using the results from \citet{ANZER2000}, the energy flux from the prominence material as a result of radiative losses is $\sim 3\times 10^4$\,erg\,cm$^{-2}$\,s$^{-1}$. Therefore, the radiative losses from a prominence thread of thickness $10^8$\,cm with the instability growing on a scale of $10^7$\,cm would approximately increase the radiative losses per unit length of the prominence thread from $2\times 10^{13}$ \,erg\,cm$^{-1}$\,s$^{-1}$ to $2.5\times 10^{13}$ \,erg\,cm$^{-1}$\,s$^{-1}$ (assuming all the losses are from optically thick lines). Though this is not as effective as the large radiative losses that occur once mixing is fully developed, it does highlight that the increase in radiative losses can occur through all stages of the instability.}

The example using prominence material embedded in the solar corona has a large temperature difference, and as such the heating possible by the KHi is limited. However, coronal loops are of a similar temperature and density to the ambient corona. In this situation, as the density contrast is small the heating will be at its most effective, and as the temperature contrast is small $T_{\rm mixed}$ will not differ greatly for the background temperatures. The latter means that the radiative losses will not be greatly affected making it possible that heating rates can outstrip loss rates, and the former means that the maximum heating would be given as $\sim M^2/16$, around the most efficient it can be.

\subsection{Some thoughts on driven oscillations}

In this section, we have focussed on the case that the KHi develops as a result of an oscillation that is driven by a impulsive kick and then left to evolve, but another possibility is that an oscillation in a flux tube in the solar atmosphere is being continuously driven at its ends in the photosphere. Assuming that the driven oscillations are not strong enough to completely destroy the structure of the flux tube, then there would be a constant energy source to drive instability and with it turbulence. This would again create the mixing layer via the KHi process presented in this paper. However, we can hypothesize that once this layer has become large enough compared to the radius of the flux tube, the boundary between the flux tube and the external corona would become sufficiently thick that instabilities cannot grow. This can be seen in our simulations (c.f. Figure 5 panel d) where even though there is still shear flow (and with it free energy that could be used for both mixing and heating) as the mixing layer has become sufficiently large compared to the width of the box it is no longer able to extract more energy from the flow. This implies that geometrical constraints on the absolute thickness of mixing layer on a flux tube exist. Inside this layer, as there are going to be radial density variations there is the possibility of resonances between the local Aflv\'{e}n frequency and the frequency of the large scale oscillation which may allow further, localised excitation of the KHi, and with that further dissipation as discussed in \citet{TERR2008}.

Another aspect to consider is the time scale of the mixing (given in Equation 33). This is limited by the velocity shear (shorter for larger shears) and by the density contrast (longer for larger contrasts), so it is easy to conceive cases where the time scale for mixing (and with it heating) is longer than the characteristic time scale for energy input into the flux tube. This can be mitigated by energy being only injected at what is relatively the small cross section of the footpoints of a flux tube, in contrast with the large regions on the flanks of a flux tube that can be dissipating energy via the KHi. This implies that even if the characteristic time scale for energy dissipation is longer, a larger region is involved in the dissipation meaning the total energy input into oscillating flows of the flux tube and the total energy extracted from these flows can balance. However, when this is not the case there are three possibilities: other dissipation mechanisms dominate, the excess energy leaks from the tube, {or} (in the case the driver is resonant with the flux tube kink frequency) the oscillations get larger and larger until the KHi heating time scale matches the energy input time scale.

\subsection{Differentiating between heating and mixing in simulations}\label{sim_comp}

When looking at the results of a 3D simulation, especially the incredibly complex simulations of the Kelvin--Helmholtz instability forming on the surface of oscillating prominence threads, it can be difficult to determine whether features that appear are as a result of heating or mixing.
Fortunately, as the mixing solution is based on having no change in internal energy, which implies conservation of pressure, it is possible to estimate the temperature distribution from mixing by just knowing the density distribution.

For the case where the initial pressure profile is constant, through application of the ideal gas law it would be expected that the temperature at any point in the mixing layer as a result of mixing alone is $T\propto1/\rho$.
For a situation where the gas pressure is not constant, this can be more complicated to estimate but it is possible to determine a linear map between the density and the expected pressure achieved through mixing.
The density at any point in the mixing layer can be written as the sum of the fractions of the densities outside the mixing layer given by:
\begin{equation}
\rho=A_1(\mathbf{x})\rho_1+A_2(\mathbf{x})\rho_2,
\end{equation}
where $A_1(\mathbf{x})+A_2(\mathbf{x})=1$, which allows for $A_1$ and $A_2$ to be uniquely determined from simulation data at each point in the mixing layer.
This means that the pressure expected from mixing at any point in the mixing layer is given by:
\begin{equation}
p=A_1(\mathbf{x})p_1+A_2(\mathbf{x})p_2.
\end{equation}
Then using an ideal gas law, the temperature at that point in the mixing layer expected to occur as a result of mixing associated with these  values of $\rho$ and $p$ can be determined.
By comparing this estimate to the simulated temperature distribution, it is possible to estimate the position and level of the temperature increase by removing the influence of the mixing.

To highlight this, we have applied these arguments to our simulations. This is presented in Figure \ref{temp_estimate}, where the difference between the local temperature of the plasma and the temperature estimated using the mixing arguments all normalised by the estimated temperature is plotted for a slice in the simulation at $z=0$. In the KHi layer there are areas where heating (either through compression or dissipation) has lead to temperature increases. However, the magnitude of these increases only reach to the level of $\sim2.5$ per cent of the temperature the plasma reaches through mixing. 

\begin{figure}
  \begin{center}
\includegraphics[height=8cm]{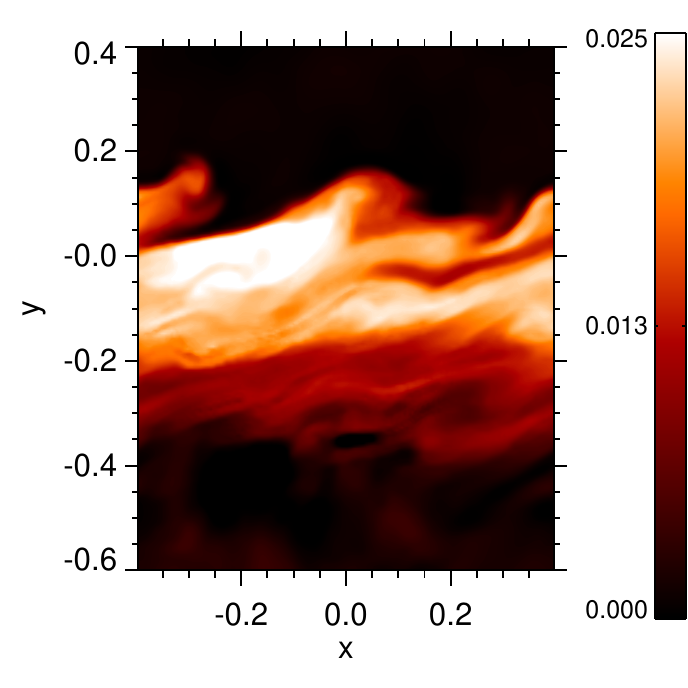}
  \end{center}
  \caption{Difference between temperature and mixing temperature normalised by mixing temperature in the KHi layer at $t=60$.}
\label{temp_estimate}
\end{figure}

There is a caveat to this estimate, and that comes as a result of compression.
The model in the previous paragraphs assumes no change in internal energy, so any change found in internal energy could be attributed to turbulent heating.
However, changes in internal energy can come about through work done through compression.
If there is an initial pressure jump or the turbulence is large enough to have a noticeable turbulent pressure there will be a compressive or expansive effect which will alter the temperature from the mixing value but without being associated with energy dissipation.
Therefore, it would always be worth estimating the order of these effects before making the comparison to make sure the heating estimate is not misleading.

\section{Summary and Discussion}

In this paper we have presented a simple phenomenological model for mixing by the magnetic Kelvin--Helmholtz instability in a uniform pressure and magnetic field.
The model was constructed using conservation of mass, momentum and energy, and this has been used to predict the characteristic values of the density, pressure, magnetic field, temperature, velocity and kinetic energy associated with the mixing layer.
The key results are:
\begin{enumerate}[(i)]
\item The central position of the mixing layer is shifted by having a density jump towards the low density side. The larger the density difference the larger the shift.

\item The characteristic density in this layer is given by $\sqrt{\rho_1\rho_2}$, and 
the characteristic value of the temperature is $T_{\rm mixed}=\sqrt{T_1T_2}$. 

\item The total fluctuating energy can be calculated and used as an estimate for the energy that can be dissipated in the system and is bounded above by $M^2/16$ where $M$ is the Mach number of the flow.

\item In high Reynolds number flows an estimate for the lower bound of the time scale for mixing/heating can be given by:
\begin{equation}
\tau_{\rm mixed}\ge\frac{2l}{\Delta V }\sqrt{2}\frac{\sqrt{\alpha_1}+\sqrt{\alpha_2}}{(\alpha_1\alpha_2)^{1/4}},
\end{equation}
with this also providing an estimate of the dissipation time scale of the system.

\item The predictions of this model are well-supported by numerical calculations.

\end{enumerate}

Application of this model to the formation of a thick transition region between a cool, high-density region and the hot, tenuous solar corona highlights that it is much more likely for the mixing to drive the observed temperature changes presented in \citet{OKA2015} than heating. 
With predictions for the temperature material created by mixing giving estimates of $10^5$\,K.
This leads to one of the greatest consequences of Kelvin--Helmholtz instability, the mixing process does not add much heat to the system, but does greatly increase the efficiency of the radiative losses {by creating thick regions at transition region temperatures and densities}.
Ultimately this process makes energy loss from the corona, i.e. coronal cooling, more likely than coronal heating. 

The current level of the model does not take into account many possible variations. For example how oscillatory flow changes the nonlinear evolution \citep[the linear stability problem was investigated in][]{HILL2019, BARB2019}, or how changes in the gas and magnetic pressure across the shear layer change the mixing process. 
However, the current model provides sufficiently accurate estimates and scalings for the basic model proposed based on the conservation of mass, momentum and energy of the system, so any extension of the model to more complex scenarios will still have the same constraints. Therefore, the general conclusions are likely not to be greatly altered.

{One area that is worthy of discussion, though beyond the scope of the current paper, is the influence of a non-potential magnetic field on the energy released. It can be expected that, as with the kinetic energy, the total of the mean magnetic energy distribution after mixing is smaller than that held in the initial distribution (see Figure \ref{KE_prof} for the change in the kinetic energy).
Simulations by \citet{HOW2017} numerically investigated this possibility, with their results suggesting that more heating would be possible as a result of the KHi developing in a twisted magnetic field.
As such, an important further development for the model we present in this paper is the inclusion of these effects.}

The model presented in this paper has been used to investigate the nonlinear MHD Kelvin--Helmholtz instability relating to solar prominences, but all the arguments in Section \ref{phem_model} are equally applicable in the case where $B=0$, i.e. it also applies to hydrodynamic systems, and to other MHD systems \citep[e.g. the flanks of CMEs][]{FOU2011}.
Therefore, though the application of the model looked at in this paper is for the solar atmosphere, in reality the model, and the extensions it promises, is significantly more versatile, and can be applied to any system undergoing the MHD KHi to estimate the mixing and heating behaviour.

\acknowledgments

AH is supported by his STFC Ernest Rutherford Fellowship grant number ST/L00397X/2 and STFC research grant ST/R000891/1. 
 IA acknowledges financial support from the Spanish Ministerio de Ciencia, Innovaci\'on y Universidades through project PGC2018-102108-B-I00 and FEDER funds.
This work used the COSMA Data Centric system at Durham University, operated by the Institute for Computational Cosmology on behalf of the STFC DiRAC HPC Facility (www.dirac.ac.uk). This equipment was funded by a BIS National E-infrastructure capital grant ST/K00042X/1, DiRAC Operations grant ST/K003267/1 and Durham University. DiRAC is part of the UK National E-Infrastructure.


\vspace{5mm}


\begin{thebibliography}{}
\expandafter\ifx\csname natexlab\endcsname\relax\def\natexlab#1{#1}\fi
\providecommand{\url}[1]{\href{#1}{#1}}
\providecommand{\dodoi}[1]{doi:~\href{http://doi.org/#1}{\nolinkurl{#1}}}
\providecommand{\doeprint}[1]{\href{http://ascl.net/#1}{\nolinkurl{http://ascl.net/#1}}}
\providecommand{\doarXiv}[1]{\href{https://arxiv.org/abs/#1}{\nolinkurl{https://arxiv.org/abs/#1}}}

\bibitem[{{Antolin} {et~al.}(2016){Antolin}, {De Moortel}, {Van Doorsselaere},
  \& {Yokoyama}}]{ANT2016}
{Antolin}, P., {De Moortel}, I., {Van Doorsselaere}, T., \& {Yokoyama}, T.
  2016, \apj, 830, L22, \dodoi{10.3847/2041-8205/830/2/L22}

\bibitem[{{Antolin} {et~al.}(2017){Antolin}, {De Moortel}, {Van Doorsselaere},
  \& {Yokoyama}}]{ANT2017}
---. 2017, \apj, 836, 219, \dodoi{10.3847/1538-4357/aa5eb2}

\bibitem[{{Antolin} {et~al.}(2015){Antolin}, {Okamoto}, {De Pontieu},
  {Uitenbroek}, {Van Doorsselaere}, \& {Yokoyama}}]{ANT2015}
{Antolin}, P., {Okamoto}, T.~J., {De Pontieu}, B., {et~al.} 2015, \apj, 809,
  72, \dodoi{10.1088/0004-637X/809/1/72}

\bibitem[{{Antolin} {et~al.}(2018){Antolin}, {Schmit}, {Pereira}, {De Pontieu},
  \& {De Moortel}}]{ANT2018}
{Antolin}, P., {Schmit}, D., {Pereira}, T.~M.~D., {De Pontieu}, B., \& {De
  Moortel}, I. 2018, \apj, 856, 44, \dodoi{10.3847/1538-4357/aab34f}

\bibitem[{{Antolin} {et~al.}(2014){Antolin}, {Yokoyama}, \& {Van
  Doorsselaere}}]{ANT2014}
{Antolin}, P., {Yokoyama}, T., \& {Van Doorsselaere}, T. 2014, \apj, 787, L22,
  \dodoi{10.1088/2041-8205/787/2/L22}

\bibitem[{{Anzer} \& {Heinzel}(2000)}]{ANZER2000}
{Anzer}, U., \& {Heinzel}, P. 2000, \aap, 358, L75

\bibitem[{{Anzer} \& {Heinzel}(2008)}]{ANZ2008}
---. 2008, \aap, 480, 537, \dodoi{10.1051/0004-6361:20078832}

\bibitem[{{Arregui}(2015)}]{ARREGUI2015}
{Arregui}, I. 2015, Philosophical Transactions of the Royal Society of London
  Series A, 373, 20140261, \dodoi{10.1098/rsta.2014.0261}

\bibitem[{{Barbulescu} {et~al.}(2019){Barbulescu}, {Ruderman}, {Van
  Doorsselaere}, \& {Erd{\'e}lyi}}]{BARB2019}
{Barbulescu}, M., {Ruderman}, M.~S., {Van Doorsselaere}, T., \& {Erd{\'e}lyi},
  R. 2019, \apj, 870, 108, \dodoi{10.3847/1538-4357/aaf506}

\bibitem[{{Berger} {et~al.}(2017){Berger}, {Hillier}, \& {Liu}}]{BERG2017}
{Berger}, T., {Hillier}, A., \& {Liu}, W. 2017, \apj, 850, 60,
  \dodoi{10.3847/1538-4357/aa95b6}

\bibitem[{{Berger} {et~al.}(2008){Berger}, {Shine}, {Slater}, {Tarbell},
  {Title}, {Okamoto}, {Ichimoto}, {Katsukawa}, {Suematsu}, {Tsuneta}, {Lites},
  \& {Shimizu}}]{BERG2008}
{Berger}, T.~E., {Shine}, R.~A., {Slater}, G.~L., {et~al.} 2008, \apjl, 676,
  L89, \dodoi{10.1086/587171}

\bibitem[{{Berger} {et~al.}(2010){Berger}, {Slater}, {Hurlburt}, {Shine},
  {Tarbell}, {Title}, {Lites}, {Okamoto}, {Ichimoto}, {Katsukawa}, {Magara},
  {Suematsu}, \& {Shimizu}}]{BERG2010}
{Berger}, T.~E., {Slater}, G., {Hurlburt}, N., {et~al.} 2010, \apj, 716, 1288,
  \dodoi{10.1088/0004-637X/716/2/1288}

\bibitem[{{Chandrasekhar}(1961)}]{CHAN1961}
{Chandrasekhar}, S. 1961, {Hydrodynamic and hydromagnetic stability}

\bibitem[{{De Pontieu} {et~al.}(2014){De Pontieu}, {Title}, {Lemen}, {Kushner},
  {Akin}, {Allard}, {Berger}, {Boerner}, {Cheung}, {Chou}, {Drake}, {Duncan},
  {Freeland}, {Heyman}, {Hoffman}, {Hurlburt}, {Lindgren}, {Mathur}, {Rehse},
  {Sabolish}, {Seguin}, {Schrijver}, {Tarbell}, {W{\"u}lser}, {Wolfson},
  {Yanari}, {Mudge}, {Nguyen-Phuc}, {Timmons}, {van Bezooijen}, {Weingrod},
  {Brookner}, {Butcher}, {Dougherty}, {Eder}, {Knagenhjelm}, {Larsen},
  {Mansir}, {Phan}, {Boyle}, {Cheimets}, {DeLuca}, {Golub}, {Gates}, {Hertz},
  {McKillop}, {Park}, {Perry}, {Podgorski}, {Reeves}, {Saar}, {Testa}, {Tian},
  {Weber}, {Dunn}, {Eccles}, {Jaeggli}, {Kankelborg}, {Mashburn}, {Pust},
  {Springer}, {Carvalho}, {Kleint}, {Marmie}, {Mazmanian}, {Pereira}, {Sawyer},
  {Strong}, {Worden}, {Carlsson}, {Hansteen}, {Leenaarts}, {Wiesmann},
  {Aloise}, {Chu}, {Bush}, {Scherrer}, {Brekke}, {Martinez-Sykora}, {Lites},
  {McIntosh}, {Uitenbroek}, {Okamoto}, {Gummin}, {Auker}, {Jerram}, {Pool}, \&
  {Waltham}}]{DEPON2014}
{De Pontieu}, B., {Title}, A.~M., {Lemen}, J.~R., {et~al.} 2014, \solphys, 289,
  2733, \dodoi{10.1007/s11207-014-0485-y}

\bibitem[{D. ~Winant  \& K. ~Browand (1974)}]{WIN1974}
D. ~Winant , C., \& K. ~Browand , F. 1974, Journal of Fluid Mechanics, 63,
  237 , \dodoi{10.1017/S0022112074001121}

\bibitem[{{Engvold} {et~al.}(1990){Engvold}, {Hirayama}, {Leroy}, {Priest}, \&
  {Tandberg-Hanssen}}]{ENG1990}
{Engvold}, O., {Hirayama}, T., {Leroy}, J.~L., {Priest}, E.~R., \&
  {Tandberg-Hanssen}, E. 1990, {Hvar Reference Atmosphere of Quiescent
  Prominences}, ed. V.~{Ruzdjak} \& E.~{Tandberg-Hanssen}, Vol. 363, 294

\bibitem[{{Foullon} {et~al.}(2011){Foullon}, {Verwichte}, {Nakariakov},
  {Nykyri}, \& {Farrugia}}]{FOU2011}
{Foullon}, C., {Verwichte}, E., {Nakariakov}, V.~M., {Nykyri}, K., \&
  {Farrugia}, C.~J. 2011, \apj, 729, L8, \dodoi{10.1088/2041-8205/729/1/L8}

\bibitem[{{Goossens} {et~al.}(2009){Goossens}, {Terradas}, {Andries},
  {Arregui}, \& {Ballester}}]{GOOS2009}
{Goossens}, M., {Terradas}, J., {Andries}, J., {Arregui}, I., \& {Ballester},
  J.~L. 2009, \aap, 503, 213, \dodoi{10.1051/0004-6361/200912399}

\bibitem[{{Hillier}(2019)}]{HILL2019b}
{Hillier}, A. 2019, arXiv e-prints, arXiv:1907.12507.
\newblock \doarXiv{1907.12507}

\bibitem[{{Hillier} {et~al.}(2019){Hillier}, {Barker}, {Arregui}, \&
  {Latter}}]{HILL2019}
{Hillier}, A., {Barker}, A., {Arregui}, I., \& {Latter}, H. 2019, \mnras, 482,
  1143, \dodoi{10.1093/mnras/sty2742}

\bibitem[{{Hillier} {et~al.}(2017){Hillier}, {Matsumoto}, \&
  {Ichimoto}}]{HILL2017}
{Hillier}, A., {Matsumoto}, T., \& {Ichimoto}, K. 2017, \aap, 597, A111,
  \dodoi{10.1051/0004-6361/201527766}

\bibitem[{{Hillier} {et~al.}(2013){Hillier}, {Morton}, \&
  {Erd{\'e}lyi}}]{HILL2013}
{Hillier}, A., {Morton}, R.~J., \& {Erd{\'e}lyi}, R. 2013, \apjl, 779, L16,
  \dodoi{10.1088/2041-8205/779/2/L16}

\bibitem[{{Hillier} \& {Polito}(2018)}]{HILL2018}
{Hillier}, A., \& {Polito}, V. 2018, \apj, 864, L10,
  \dodoi{10.3847/2041-8213/aad9a5}

\bibitem[{{Hillier} {et~al.}(2016){Hillier}, {Takasao}, \&
  {Nakamura}}]{HILL2016}
{Hillier}, A., {Takasao}, S., \& {Nakamura}, N. 2016, \aap, 591, A112,
  \dodoi{10.1051/0004-6361/201628215}

\bibitem[{{Hollweg}(1978)}]{HOLL1978}
{Hollweg}, J.~V. 1978, \solphys, 56, 305, \dodoi{10.1007/BF00152474}

\bibitem[{{Howson} {et~al.}(2017){Howson}, {De Moortel}, \&
  {Antolin}}]{HOW2017}
{Howson}, T.~A., {De Moortel}, I., \& {Antolin}, P. 2017, \aap, 607, A77,
  \dodoi{10.1051/0004-6361/201731178}

\bibitem[{{Ionson}(1978)}]{ION1978}
{Ionson}, J.~A. 1978, \apj, 226, 650, \dodoi{10.1086/156648}

\bibitem[{{Karampelas} {et~al.}(2017){Karampelas}, {Van Doorsselaere}, \&
  {Antolin}}]{KARA2017}
{Karampelas}, K., {Van Doorsselaere}, T., \& {Antolin}, P. 2017, \aap, 604,
  A130, \dodoi{10.1051/0004-6361/201730598}

\bibitem[{{Kelvin}(1880)}]{KEL1880}
{Kelvin}, L. 1880, Nature, 23, 45, \dodoi{10.1038/023045a0}

\bibitem[{{Labrosse} {et~al.}(2010){Labrosse}, {Heinzel}, {Vial}, {Kucera},
  {Parenti}, {Gun{\'a}r}, {Schmieder}, \& {Kilper}}]{LAB2010}
{Labrosse}, N., {Heinzel}, P., {Vial}, J.~C., {et~al.} 2010, \ssr, 151, 243,
  \dodoi{10.1007/s11214-010-9630-6}

\bibitem[{{Magyar} \& {Van Doorsselaere}(2016)}]{MAGYAR2016}
{Magyar}, N., \& {Van Doorsselaere}, T. 2016, \aap, 595, A81,
  \dodoi{10.1051/0004-6361/201629010}

\bibitem[{{Mak} {et~al.}(2017){Mak}, {Griffiths}, \& {Hughes}}]{MAK2017}
{Mak}, J., {Griffiths}, S.~D., \& {Hughes}, D.~W. 2017, Physical Review Fluids,
  2, 113701, \dodoi{10.1103/PhysRevFluids.2.113701}

\bibitem[{{Matsumoto} \& {Seki}(2010)}]{MATSU2010}
{Matsumoto}, Y., \& {Seki}, K. 2010, Journal of Geophysical Research (Space
  Physics), 115, A10231, \dodoi{10.1029/2009JA014637}

\bibitem[{{M{\"o}stl} {et~al.}(2013){M{\"o}stl}, {Temmer}, \&
  {Veronig}}]{MOSTL2013}
{M{\"o}stl}, U.~V., {Temmer}, M., \& {Veronig}, A.~M. 2013, \apj, 766, L12,
  \dodoi{10.1088/2041-8205/766/1/L12}

\bibitem[{{Ofman} \& {Thompson}(2011)}]{OFMAN2011}
{Ofman}, L., \& {Thompson}, B.~J. 2011, \apj, 734, L11,
  \dodoi{10.1088/2041-8205/734/1/L11}

\bibitem[{{Okamoto} {et~al.}(2015){Okamoto}, {Antolin}, {De Pontieu},
  {Uitenbroek}, {Van Doorsselaere}, \& {Yokoyama}}]{OKA2015}
{Okamoto}, T.~J., {Antolin}, P., {De Pontieu}, B., {et~al.} 2015, \apj, 809,
  71, \dodoi{10.1088/0004-637X/809/1/71}

\bibitem[{{Okamoto} {et~al.}(2007){Okamoto}, {Tsuneta}, {Berger}, {Ichimoto},
  {Katsukawa}, {Lites}, {Nagata}, {Shibata}, {Shimizu}, {Shine}, {Suematsu},
  {Tarbell}, \& {Title}}]{OKA2007}
{Okamoto}, T.~J., {Tsuneta}, S., {Berger}, T.~E., {et~al.} 2007, Science, 318,
  1577, \dodoi{10.1126/science.1145447}

\bibitem[{{Patsourakos} \& {Vial}(2002)}]{PATS2002}
{Patsourakos}, S., \& {Vial}, J.-C. 2002, \solphys, 208, 253,
  \dodoi{10.1023/A:1020510120772}

\bibitem[{{Rempel} {et~al.}(2009){Rempel}, {Sch{\"u}ssler}, \&
  {Kn{\"o}lker}}]{REM2009}
{Rempel}, M., {Sch{\"u}ssler}, M., \& {Kn{\"o}lker}, M. 2009, \apj, 691, 640,
  \dodoi{10.1088/0004-637X/691/1/640}

\bibitem[{{Ryu} {et~al.}(2000){Ryu}, {Jones}, \& {Frank}}]{RYU2000}
{Ryu}, D., {Jones}, T.~W., \& {Frank}, A. 2000, \apj, 545, 475,
  \dodoi{10.1086/317789}

\bibitem[{{Ryutova} {et~al.}(2010){Ryutova}, {Berger}, {Frank}, {Tarbell}, \&
  {Title}}]{RYU2010}
{Ryutova}, M., {Berger}, T., {Frank}, Z., {Tarbell}, T., \& {Title}, A. 2010,
  \solphys, 267, 75, \dodoi{10.1007/s11207-010-9638-9}

\bibitem[{{Soler} {et~al.}(2010){Soler}, {Terradas}, {Oliver}, {Ballester}, \&
  {Goossens}}]{SOLER2010}
{Soler}, R., {Terradas}, J., {Oliver}, R., {Ballester}, J.~L., \& {Goossens},
  M. 2010, \apj, 712, 875, \dodoi{10.1088/0004-637X/712/2/875}

\bibitem[{{Stuart}(1967)}]{STU1967}
{Stuart}, J.~T. 1967, Journal of Fluid Mechanics, 29, 417,
  \dodoi{10.1017/S0022112067000941}

\bibitem[{{Terradas} {et~al.}(2008){Terradas}, {Andries}, {Goossens},
  {Arregui}, {Oliver}, \& {Ballester}}]{TERR2008}
{Terradas}, J., {Andries}, J., {Goossens}, M., {et~al.} 2008, \apj, 687, L115,
  \dodoi{10.1086/593203}

\bibitem[{{Terradas} {et~al.}(2018){Terradas}, {Magyar}, \& {Van
  Doorsselaere}}]{TERR2018}
{Terradas}, J., {Magyar}, N., \& {Van Doorsselaere}, T. 2018, \apj, 853, 35,
  \dodoi{10.3847/1538-4357/aa9d0f}

\bibitem[{{Tsuneta} {et~al.}(2008){Tsuneta}, {Ichimoto}, {Katsukawa}, {Nagata},
  {Otsubo}, {Shimizu}, {Suematsu}, {Nakagiri}, {Noguchi}, {Tarbell}, {Title},
  {Shine}, {Rosenberg}, {Hoffmann}, {Jurcevich}, {Kushner}, {Levay}, {Lites},
  {Elmore}, {Matsushita}, {Kawaguchi}, {Saito}, {Mikami}, {Hill}, \&
  {Owens}}]{TSU2008}
{Tsuneta}, S., {Ichimoto}, K., {Katsukawa}, Y., {et~al.} 2008, \solphys, 249,
  167, \dodoi{10.1007/s11207-008-9174-z}

\bibitem[{Vallis(2017)}]{VAL2017}
Vallis, G.~K. 2017, Atmospheric and Oceanic Fluid Dynamics: Fundamentals and
  Large-Scale Circulation, 2nd edn. (Cambridge, U.K.: Cambridge University
  Press), 946

\bibitem[{{Wentzel}(1974)}]{WEN1974}
{Wentzel}, D.~G. 1974, \solphys, 39, 129, \dodoi{10.1007/BF00154975}

\bibitem[{{Wentzel}(1978)}]{WEN1978}
---. 1978, Reviews of Geophysics and Space Physics, 16, 757,
  \dodoi{10.1029/RG016i004p00757}

\bibitem[{{Wentzel}(1979)}]{WEN1979}
---. 1979, \apj, 233, 756, \dodoi{10.1086/157437}

\bibitem[{{Yang} {et~al.}(2018){Yang}, {Xu}, {Lim}, {Kim}, {Cho}, {Kim},
  {Chae}, {Cho}, \& {Ji}}]{YANG2018}
{Yang}, H., {Xu}, Z., {Lim}, E.-K., {et~al.} 2018, \apj, 857, 115,
  \dodoi{10.3847/1538-4357/aab789}

\end{thebibliography}
\end{document}